\documentclass[twoside]{article}
\usepackage{qic}
\usepackage{epsfig} 

\let\square\relax 
\usepackage{amsmath}
\usepackage{amssymb}
\usepackage{latexsym, color, hyperref}
\usepackage{authblk}
\usepackage{wasysym} 
\usepackage{enumerate}
\usepackage{kbordermatrix}
\usepackage{graphicx}
\usepackage{algorithm}
\usepackage[retainorgcmds]{IEEEtrantools}
\allowdisplaybreaks[1]

\begin{document}
\setlength{\textheight}{8.0truein}    

\runninghead{Efficient optimization of perturbative gadgets}
            {Cao, Y. and Kais, S.}

\normalsize\textlineskip
\thispagestyle{empty}
\setcounter{page}{1}

\copyrightheading{17}{9 \& 10}{2017}{0779--0809}

\vspace*{0.88truein}

\alphfootnote

\fpage{1}

\newtheorem{claim}{Claim}
\newtheorem{remark}{Remark}

\newcommand{\note}[1]{{\color{red} #1}}
\newcommand{\yudong}[1]{{\color{blue}#1}}
\newcommand{\ii}{\mathbb{I}}
\newcommand{\CNOT}{\mathrm{CNOT}}
\newcommand{\defeq}{\stackrel{\mathrm{def}}{=}}
\newcommand{\poly}{\mathrm{poly}}
\newcommand{\const}{\mathrm{const}}
\newcommand{\prob}[1]{\Pr[ #1 ]}
\newcommand{\norm}[1]{\left\| #1 \right\|}
\newcommand{\oo}[1]{\Theta\left(#1\right)} 
\newcommand{\bra}[1]{\langle #1 \vert}
\newcommand{\braL}[1]{\left\langle #1 \right\vert}
\newcommand{\ket}[1]{\vert #1 \rangle}
\newcommand{\ketL}[1]{\left\vert #1 \right\rangle}
\newcommand{\ketbra}[1]{\vert #1 \rangle \langle #1 \vert}
\newcommand{\braket}[2]{\langle #1 \vert #2 \rangle}
\newcommand{\conjugate}[1]{#1^{\dagger}}
\newcommand{\astconjugate}[1]{#1^{\ast}}
\newcommand{\complexconjugate}[1]{\overline{#1}}
\newcommand{\transpose}[1]{#1^{\mathsf{T}}}
\newcommand{\rootwo}{\frac{1}{\sqrt{2}}}
\newcommand{\half}{\frac{1}{2}}
\newcommand{\nono}{{\em no }}
\newcommand{\yeye}{{\em yes }}
\newcommand{\vvv}[2]{\left[ \begin{array}{c} #1 \\ #2 \end{array}\right]}
\newcommand{\vww}[4]{\left[ \begin{array}{c} #1 \\ #2 \\ #3 \\ #4\end{array}\right]}
\newcommand{\mmm}[4]{\left[ \begin{array}{cc} #1 & #2 \\ #3 & #4\end{array}\right]}

\renewcommand{\sfdefault}{phv}
\renewcommand{\rmdefault}{ptm}
\newlength{\actualtopmargin}
\newlength{\actualsidemargin}
\setlength{\actualtopmargin}{2cm}
\setlength{\actualsidemargin}{2.0cm}
\setlength{\topmargin}{-1.0in}
  \addtolength{\topmargin}{-\headsep}
  \addtolength{\topmargin}{-\headheight}
  \addtolength{\topmargin}{\actualtopmargin}
\addtolength{\oddsidemargin}{-\evensidemargin}
  \setlength{\oddsidemargin}{0.35\oddsidemargin}
  \addtolength{\oddsidemargin}{\actualsidemargin}
  \addtolength{\oddsidemargin}{-1.0in}
\setlength{\evensidemargin}{-\oddsidemargin}
  \addtolength{\evensidemargin}{2\actualsidemargin}
  \addtolength{\evensidemargin}{-2.0in}
\setlength{\textheight}{\paperheight}
  \addtolength{\textheight}{-2\actualtopmargin}
\setlength{\textwidth}{\paperwidth}
  \addtolength{\textwidth}{-2\actualsidemargin}

\centerline{\bf
Efficient optimization of perturbative gadgets}
\vspace*{0.37truein}
\centerline{\footnotesize
YUDONG CAO}
\vspace*{0.015truein}
\centerline{\footnotesize\it Department of Computer Science, Purdue University, West Lafayette, IN 47906, USA}
\baselineskip=10pt
\centerline{\footnotesize\it Department of Chemistry and Chemical Biology, Harvard University, Cambridge, MA 02138}
\vspace*{10pt}
\centerline{\footnotesize 
SABRE KAIS}
\vspace*{0.015truein}
\centerline{\footnotesize\it Department of Chemistry, Purdue University, West Lafayette, IN 47906, USA}
\baselineskip=10pt
\centerline{\footnotesize\it Qatar Energy and Environment Research Institute, HBKU, Doha, Qatar}
\baselineskip=10pt
\centerline{\footnotesize\it Santa Fe Institute, 1399 Hyde Park Rd, Santa Fe, NM 87501, USA}
\vspace*{0.225truein}
\publisher{November 5, 2016}{June 6, 2017}

\vspace*{0.21truein}

 
\abstract{
Perturbative gadgets are general techniques for reducing many-body spin interactions to two-body ones using perturbation theory. This allows for potential realization of effective many-body interactions using more physically viable two-body ones. In parallel with prior work (arXiv:1311.2555 [quant-ph]), here we consider minimizing the physical resource required for implementing the gadgets initially proposed by Kempe, Kitaev and Regev (arXiv:quant-ph/0406180) and later generalized by Jordan and Farhi (arXiv:0802.1874v4). The main innovation of our result is a set of methods that efficiently compute tight upper bounds to errors in the perturbation theory. We show that in cases where the terms in the target Hamiltonian commute, the bounds produced by our algorithm are sharp for arbitrary order perturbation theory. We provide numerics which show orders of magnitudes improvement over gadget constructions based on trivial upper bounds for the error term in the perturbation series. We also discuss further improvement of our result by adopting the Schrieffer-Wolff formalism of perturbation theory and supplement our observation with numerical results.
}

\vspace*{10pt}

\keywords{Quantum many-body problem, perturbation theory, spin systems}
\vspace*{3pt}
\communicate{Barbara Terhal and Richard Jozsa}

\vspace*{1pt}\textlineskip    


\section{Introduction}

Quantum many-body interactions arise in a variety of contexts in quantum information and quantum computation, such as topological quantum computing \cite{Kitaev20032,RevModPhys.80.1083,quant-ph/0101025,Ogburn:1998:TQC:645812.670805}, measurement-based model of quantum computing \cite{PhysRevLett.86.5188,PhysRevA.68.022312,PhysRevLett.86.910,PhysRevA.71.062313,quant-ph/0603226,doi:10.1080/09500340110107487}, adiabatic simulation of quantum chemistry \cite{BLA14}, universal adiabatic quantum computation\footnote{We note that there are also several proposals \cite{PhysRevLett.99.070502,1367-2630-18-2-023042,PhysRevLett.114.140501} of universal adiabatic quantum computation that uses only simple two-body interactions, thus circumventing the need for many-body interactions. On the other hand, some of these proposals such as \cite{1367-2630-18-2-023042} also involve perturbation theory for which the error estimation algorithms of this paper may be useful for optimizing the parameters of these constructions as well.} \cite{ADKL+07,PhysRevA.75.062337}, as well as constructions of circuit-to-Hamiltonian mapping for QMA-completeness \cite{KSV02,KKR06}. Given the broad range of applications for many-body interactions, it is then of great interest to simulate the behaviours of these many-body systems using experimental quantum systems. However, the current technologies for realizing controllable quantum interactions are limited to two-body interactions, implying a need for reducing many-body interactions to two-body ones. Such reduction boils down to constructing a two-body Hamiltonian whose low-lying eigenspace captures the eigenvectors and eigenvalues of the many-body ``target" Hamiltonian. The technique of \emph{perturbative gadgets} \cite{KKR06,OT06,BDLT08,JF08} fulfills precisely this task. 

The basic idea of perturbative gadget is that given a many-body ``target" Hamiltonian $H_\text{targ}$, we construct a two-body ``gadget" Hamiltonian $\tilde{H}$ of the form $H+V$ such that the low energy effective Hamiltonian of $\tilde{H}$ is arbitrarily close to $H_\text{targ}$. The gadget Hamiltonian $\tilde{H}$ acts on not only the Hilbert space of the target Hamiltonian but also an ancilla space. In other words we are embedding the spectrum of a given many-body Hamiltonian onto the low energy sector of a two-body Hamiltonian that acts on a larger Hilbert space. The effectiveness of such embedding is established by using perturbation theory for computing the low-energy effective interaction of $\tilde{H}$ and show that terms involving $H_\text{targ}$ appear at the first few orders and the total contribution from the remaining terms in the infinite series amounts to a small quantity.

As useful as the perturbative gadgets have been in the study of the complexity of various types of physical systems \cite{KKR06,OT06,BDOT06,Cubitt:2014:CCL:2706700.2707456}, the need for convergence in the perturbation series requires high variability in the coupling strengths that appear in the gadget Hamiltonian \cite{CRBK14}, which impose challenges for experimentally implementating the gadget Hamiltonians. Constructions that avoid using perturbation theory for reducing from many-body to two-body interactions have indeed been proposed \cite{B08,OY11,PhysRevA.94.012342}. However, as far as the authors are aware of, most of the non-perturbative constructions can be applied on general many-body Hamiltonians in the same way as their perturbative counterparts, in the sense that the non-perturbative constructions always assume that the Hamiltonian of the entire system must take certain form, while perturbative gadgets can be applied to reduce any subset of terms in a target Hamiltonian to two-body without concerning the form of the other terms in the Hamiltonian. A possible exception is perhaps a recent numerical optimization approach for finding many-body to two-body reductions \cite{BMS+17}. However, it is unclear how the cost of performing such optimization scales as the number of qubits and the number of $k$-local terms in the target Hamiltonian.

Here we consider minimizing variability in coupling strengths in the gadget Hamiltonians. This is important because it directly translates to reducing the physical resource required for experimentally implementing perturbative gadgets. Prior efforts \cite{CRBK14} have optimized gadget constructions in \cite{OT06, BDLT08} for reducing many-body interactions to two-body. Here we are interested in the gadget construction due to Kempe, Kitaev, Regev \cite{KKR06} and later generalized by Jordan and Farhi \cite{JF08}. The perturbative analysis of this construction is significantly more involved than constructions analyzed previously in \cite{CRBK14}. However, it is of interest to us due to numerical evidence in \cite{BLA14} using direct diagonalization of target and gadget Hamiltonians which suggests that this construction requires less variable range of couplings than the constructions presented in \cite{OT06, BDLT08}. 

The technique for optimizing the gadgets presented in this work generalizes our previous work \cite{CRBK14} for the gadgets in \cite{OT06, BDLT08} and applies the general framework presented in \cite{CK16}. The main innovation of our result is an efficient method for finding tight upper bounds for the error in the perturbation series (\emph{i.e.\ }the sum of terms from a specific finite order to infinity). By ``efficient" we mean that suppose the gadget Hamiltonian acts on $n$ qubits, our algorithm finds a tight upper bound and sometimes the \emph{exact} expression for perturbation terms at any order $r$ in time $O(n^r)$, even though each term in the perturbative expansion is of dimension $O(2^n)$. Of course, the efficiency of our method heavily exploits the structure of the gadget Hamiltonian \cite{KKR06,JF08} and does not necessarily hold for general perturbation theory on spin systems. However, in \cite{CK16,caothesis} we argue that the assumptions needed for establishing efficiency may apply for a broader class of physical Hamiltonians than perturbative gadgets.

\section{Perturbation theory and perturbative gadgets}\label{sec:pt}

The basic setting that we consider for perturbative analysis is a Hamiltonian $\tilde{H}=H+V$ where $H$ is diagonal in the computational basis with an energy gap $\Delta$ between the ground space and the first excited space, and $V$ is a perturbation that contains some non-zero off-diagonal elements. The main formalism that we use for extracting the low-energy effective Hamiltonian of $\tilde{H}$ is the well-known Feynman-Dyson series \cite{FW71} based on self energy. There are various other formulations of perturbation theory such as Schrieffer-Wolff transformation \cite{BDL11,PhysRev.149.491}, Bloch expansion \cite{BLOCH1975481} and Rayleigh-Schr\"{o}dinger perturbation theory (see for example \cite[Ch.\ 17]{Shankar94}). However, in the context of present work we focus on self-energy expansion from Feynman-Dyson series. In Section \ref{sec:sw} we will apply Schrieffer-Wolff transformation onto the gadget Hamiltonians and show a connection between Feynman-Dyson series and Schrieffer-Wolff transformation.

Define the subspace spanned by eigenstates of $H$ with energy lower than $\Delta/2$ as the \emph{low-energy subspace} $\mathcal{L}_-$ and its orthogonal complement as the \emph{high-energy subspace} $\mathcal{L}_+$. The projectors onto these subspaces are defined as $\Pi_-$ and $\Pi_+$ respectively. We then introduce the notation for projections of any operator $O$ onto the subspaces: $O_+\equiv\Pi_+O\Pi_+$, $O_-\equiv\Pi_-O\Pi_-$. $O_{-+}\equiv\Pi_-O\Pi_+$ and $O_{+-}\equiv\Pi_+O\Pi_-$. The setup of self-energy expansion $\Sigma_-(z)$ requires the definition of operator valued resolvents $G(z)=(zI-H)^{-1}$ and $\tilde{G}(z)=(zI-\tilde{H})^{-1}$ where $z$ is a scalar and $I$ is the identity matrix. Then the expression for self energy can be written as $\Sigma_-(z)=zI_--[\tilde{G}_-(z)]^{-1}$. Using Taylor expansion we have \cite{KKR06}
\begin{equation}\label{eq:sigmaz}
\begin{array}{ccl}
\Sigma_-(z) & = & H_- + V_- + V_{-+}G_+V_{+-} + V_{-+}G_+V_+G_+V_{+-} + \cdots \\
& = & H_0 + V_- + T_2 + T_3 + \cdots.
\end{array}
\end{equation}
Note that the $r$-th order term is simply a matrix product 
\begin{equation}\label{eq:T_r}
T_r=V_{-+}G_+(V_+G_+)^{r-2}G_+V_{+-}, 
\end{equation}
which gives rise to our later discussion (Section \ref{sec:errorbound}) that interprets it as a sum of walks on a graph. The self energy expansion is useful because it approximates the low-energy sector of the perturbed Hamiltonian $\tilde{H}$. This is captured precisely in \cite[Theorem 3]{KKR06}, which we restate below. In the present paper our analyses involve mainly two types of operator norms, namely the 2-norm $\|A\|_2=\max_{|\psi\rangle}\|A|\psi\rangle\|_2$, which is equal to the ``spectral radius" for Hermitian operators, and $\infty$-norm $\|A\|_\infty=\max_{i}\sum_j|\langle i|A|j\rangle|$, the ``maximum row sum", where $\{|i\rangle\}$ is the set of computational basis states.

\begin{theorem}[\cite{KKR06}, Theorem 3 restated]\label{thm:KKR06}
Given a Hamiltonian $\tilde H=H+V$ with $H$ having a spectral gap $\Delta$ between the ground space and the first excited subspace, suppose $\|V\|_2\le\Delta/2$. If there exists a Hamiltonian $H_\text{eff}$ whose energies are contained in the interval $[a,b]$ and some real constant $\epsilon>0$ such that $a<b<\Delta/2-\epsilon$ and for $z\in[a-\epsilon,b+\epsilon]$, we have \[\|\Sigma_-(z)-H_\text{eff}\|_2\le\epsilon,\] then the $j$-th eigenvalue $\tilde\lambda_j$ of $\tilde{H}_-$ and the corresponding $j$-th eigenvalue of $H_\text{eff}$ differ by at most $\epsilon$, for any appropriate range of $j$ values.
\end{theorem}

Theorem \ref{thm:KKR06} states that closeness in the operator norm between the self energy $\Sigma_-(z)$ and $H_\text{eff}$ implies closeness in eigenvalues. In fact it also implies closeness in eigenvectors (see \cite[Lemma 11]{KKR06}). Hence the entire operator $H_\text{eff}$ is captured by the low energy sector of $\tilde{H}$. The basic idea of perturbative gadgets is that for a given many-body Hamiltonian $H_\text{targ}$, one constructs a two-body Hamiltonian $\tilde{H}=H+V$ with $H$ and $V$ matching the setting described before Equation \ref{eq:sigmaz}, and the self energy expansion according to Equation \ref{eq:sigmaz} contains $H_\text{targ}$ in its leading orders which constitutes the effective Hamiltonian $H_\text{eff}$, while the remaining terms can be bounded from above by $\epsilon$. The gadget construction \cite{KKR06,JF08} considered in this work reduces an arbitrary many-body Hamiltonian 
\begin{equation}\label{eq:htarg}
H_\text{targ}=\sum_{i=1}^mc_iH_{\text{targ},i}
\end{equation}
where each 
$
H_{\text{targ},i}=\sigma_{i,1}\sigma_{i,2}\cdots\sigma_{i,k}
$
is a $k$-body term with $\sigma_{i,j}\in\{X,Y,Z,I\}$ being the $j$-th single-qubit operator in the $i$-th term in $H_\text{targ}$, to two-body. Here $X$, $Y$, and $Z$ are Pauli operators and $I$ is the identity operator. The gadget Hamiltonian $\tilde{H}=H+V$ works by first introducing a register of $k$ ancilla qubits for each $k$-body term $H_{\text{targ},i}$. Hence there are $km$ ancillas in total, each of which can be labelled as $(i,j)$ with $i\in[m]$ (we use $[x]$ to denote the set $\{1,\cdots,x\}$) being the index of the register and $j\in[k]$ being the index of the ancilla within the register. For each register $i$ we then impose a Hamiltonian $H^{(i)}$ which ferromagnetically couples every pair of qubits in the register. Precisely, $H=\sum_{i=1}^mH^{(i)}$ is defined with each $H^{(i)}$ having the form
\begin{equation}\label{eq:Hi}
H^{(i)} = \sum_{1\le s<t\le k}\frac{\Delta}{2(k-1)}(I-Z_{i,s}Z_{i,t}).
\end{equation}
Here in Equation \ref{eq:Hi} the operator $Z_{i,j}$ acts on the ancilla qubit $(i,j)$. Accordingly, the perturbation $V = \sum_{i=1}^mV^{(i)}$ consists of the terms that couple each register of ancillas with the corresponding qubits that the $\sigma_{i,j}$ terms act on:
\begin{equation}\label{eq:Vi}
V^{(i)} = \sum_{j=1}^k\lambda_{i,j}\sigma_{i,j}\otimes X_{i,j}.
\end{equation}
Note that the gadget Hamiltonian considered here is slightly different from the original constructions \cite{KKR06,JF08} in that the spectral gap $\Delta$ is introduced in $H^{(i)}$ in Equation \ref{eq:Hi} and $\lambda_{i,j}$ are coupling coefficients that are assigned such that the effective low energy Hamiltonian of $\tilde H$ calculated using perturbation theory in Equation \ref{eq:sigmaz} gives rise to the target Hamiltonian in Equation \ref{eq:htarg}.

The spectrum of each $H^{(i)}$ is easy to find: the subspace of states with $j$ qubits in $|1\rangle$ state has energy $E_j=\frac{j(k-j)}{k-1}\Delta$. The ground state subspace of each register of ancillas is $\mathcal{L}_{-}^{(i)}=\text{span}\{|0\rangle^{\otimes k},|1\rangle^{\otimes k}\}$. The gap between the ground state subspace and the first excited subspace is $\Delta$. The perturbation terms $V^{(i)}$ break the degeneracy of $\mathcal{L}_{-}^{(i)}$ and are set up such that the perturbed subspace approximates the spectrum of $H_\text{eff}$ closely. If one applies the self energy expansion (Equation \ref{eq:sigmaz}) to the gadget Hamiltonian defined according to Equations \ref{eq:Hi} and \ref{eq:Vi}, it is apparent that at any order $r\le k$, $T_r$ is proportional to projection $\Pi_-$ onto $\mathcal{L}_-$, since the only $r$-step transitions under $V$ that non-trivially contribute to $T_r$ are the ones that start from $|0\rangle^{\otimes k}$ (resp.\ $|1\rangle^{\otimes k}$) and return to  $|0\rangle^{\otimes k}$ (resp.\ $|1\rangle^{\otimes k}$). At the $k$-th order, if $k$ is odd, then $T_k$ consists of only a linear combination of $H_{\text{targ},i}$ terms, since only $k$-step transitions that goes from $|0\rangle^{\otimes k}$ (resp.\ $|1\rangle^{\otimes k}$) to $|0\rangle^{\otimes k}$ (resp.\ $|1\rangle^{\otimes k}$). If $k$ is even, then $T_k$ consists of terms proportional to $H_{\text{targ},i}$ as well as a term proportional to $\Pi_-$. Substituting the $H$ and $V$ in Equations \ref{eq:Hi} and \ref{eq:Vi} into Equation \ref{eq:sigmaz} leads to a self energy of the form
\begin{equation}\label{eq:sigmaz_heff}
\Sigma_-(z)=\underbrace{\gamma \Pi_- + \sum_{i=1}^m H_{\text{targ},i}\otimes \Pi_{X,i}}_{H_\text{eff}} + T_{k+1} + T_{k+2} + \cdots
\end{equation}
where $\gamma=\gamma_1+\cdots+\gamma_k$ is a scalar which sums over all contributions up to the $k$-th order and $\Pi_{X,i}=|0\rangle\langle 1|^{\otimes k}+|1\rangle\langle 0|^{\otimes k}$ acts on the $i$-th register, where we will show that each $\gamma_i$ is efficiently computable by Algorithm \ref{alg:perturbbound} in Section \ref{subsec:algo}. The remaining terms $T_{k+1}$ and so on are error terms that should be suppressed to below $\epsilon$ by assigning $\Delta$ to be appropriately large. 

\section{Improving and optimizing gadget constructions}\label{sec:improve}

\begin{figure}
\begin{center}
\includegraphics[scale=0.8]{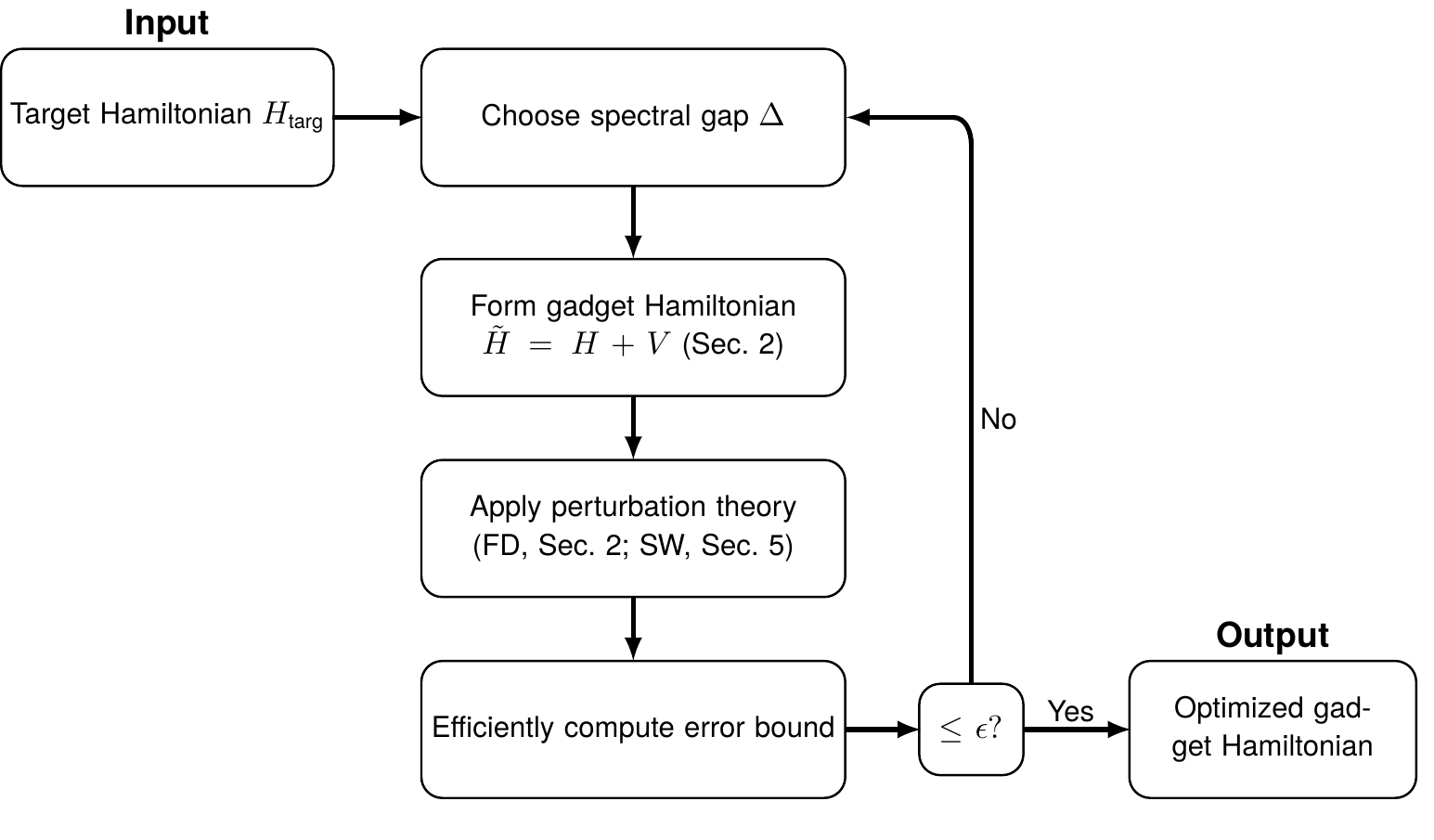}
\fcaption{The flow chart of optimizing the gadget Hamiltonian considered in this paper. We start by choosing a value $\Delta$ based on simple but loose error bounds that are dependent on $\epsilon$, such as Equation \ref{eq:Tr}. Then we construct the gadget Hamiltonian and use perturbation theory to find the effective Hamiltonian up to a certain order (FD stands for Feynman-Dyson series, which is introduced in Section \ref{sec:pt} and SW stands for Schrieffer-Wolff transformation, which will be introduced in Section \ref{sec:sw}). The norm of terms from the order of Hamiltonian on to infinite order are bounded from above by efficiently computed error bounds. Techniques for producing tight error bounds without extensive computation is the central theme of the present paper.}
\label{fig:flowchart}
\end{center}
\end{figure}


From the construction presented in the previous section, note that here the energy gap $\Delta$ is a crucial parameter that decides how accurate the perturbation theory is when applied to the gadget Hamiltonian $H+V$. The larger $\Delta$ is, the more accurately the gadget Hamiltonian captures the spectrum of the target Hamiltonian in its low energy subspace. However, larger values of $\Delta$ means more challenges for realizing the gadget Hamiltonian on an experimental system. This is because realizing the gadget Hamiltonian on a physical quantum system requires setting the coupling strengths of both the unperturbed Hamiltonian $H$, which is of the magnitude of $\Delta$, and that of the perturbation $V$, whose strength could differ substantially from that of $H$. In other words, the requirement for variability in coupling strength becomes more stringent as $\Delta$ increases. For a system of $n$ qubits with poly$(n)$ many-body terms, typically $\Delta$ scales as poly$(n)$ \cite{OT06}, which is unphysical for physical systems whose interactions are local.

The scaling of $\Delta$ as poly$(n)$ is one of the main reasons for hesitation among researchers in using perturbative gadgets for reducing many-body interactions to two-body ones, thus motivating various non-perturbative constructions for special cases \cite{B08,OY11,PhysRevA.94.012342}. This situation can be remedied for perturbative gadgets by either insisting on assigning $\Delta$ independent of system size at a cost of \emph{extensive} error $O(n\epsilon)$ \cite{BDLT08}, or substituting the current gadget construction with one that requires \emph{arbitrarily weak} interaction strengths at a cost of poly$(\epsilon^{-1})$ qubits \cite{CN14}. Whichever gadget construction one wishes to adopt, there is always a practical optimization question: what is the \emph{minimum} value of $\Delta$ such that the error does not exceed $\epsilon$? In other words, what is the value of $\Delta$ for which the error $\|\Sigma_-(z)-H_\text{eff}\|_2$ is precisely $\epsilon$? This is the question that we address in this paper, as minimizing $\Delta$ is essentially minimizing the physical resource needed for realizing the gadget Hamiltonian.

In Figure \ref{fig:flowchart} we present a flow chart of this optimization process. A challenge one has to face is then how to compute the error $\|\Sigma_-(z)-H_\text{eff}\|_2$, which is generally hard (on a classical computer) simply due to the exponential size of the Hilbert space as system size grows. We pursue a different strategy, which is to seek an upper bound of the form $\|\Sigma_-(z)-H_\text{eff}\|_2\le\|T_{k+1}\|_2+\|T_{k+2}\|_2+\cdots$, without requiring exponential-size computation in the number of qubits. In \cite{CRBK14} we use the upper bound 
\begin{equation}\label{eq:Tr}
\|T_r\|_2=\|V_{-+}(G_+V_+)^{r-2}G_+V_{+-}\|_2\le\frac{1}{|z-\Delta|^{r-1}}\|V_{-+}\|_2^2\cdot\|V_+\|_2^{r-2}
\end{equation}
which appears to be tight \cite[Figures 2 and 4]{CRBK14} for the case where $m=1$ \emph{i.e.\ }there is only one target term to be reduced and one is using gadget constructions from \cite{OT06}.                                                                                                                                                                                                                                                                                                                                                                                                                                                                                                                                                                                                                                                                                                                                                                                                                                                                                                                                                                                                                                                                                                                                                                                                                                                                                                                                                                                                                                                                                                                                                                                                                                                                                                                                                                                                                                                                                                                                                                                                                                                                                                                                                                                                                                                                                                                                                                                                                                                                                                                                                                                                                                                                                                                                                                                                                                              However, when $m>1$ and multiple gadgets are applied, the upper bound in Equation \ref{eq:Tr} becomes loose \cite[Figure 3b]{CRBK14}. This is because when perturbation theory of multiple ancilla qubits are concerned, the crude upper bound in Equation \ref{eq:Tr} could no longer capture the fine-grained details of the matrix product involved in $T_r$. Therefore we are also unable to use Equation \ref{eq:Tr} for finding tight error bounds for the gadget constructions presented in Section \ref{sec:pt}, since we are dealing with registers of ancillas of size $k>1$. 

We will show that by considering the more fine-grained details of matrix multiplication in $T_r$ and exploiting the structure of the construction in Equations \ref{eq:Hi} and \ref{eq:Vi}, it is possible to find a tight upper bound to $\|T_r\|_2$ in $O(m^r)$ time, which is polynomial in $m$ for fixed $r$, thus enabling efficient optimization of $\Delta$ in the gadget construction. 

As a final remark of the section, we note that it suffices to consider $T_r$ for fixed $r$ even though in the perturbation series $r$ goes to infinity. Let $\lambda=\max_{i,j}\lambda_{i,j}$ with $\lambda_{i,j}$ defined in Equation \ref{eq:Vi}. Our goal for the gadget construction is that at $k$-th order we have the target Hamiltonian in Equation \ref{eq:htarg} with coupling coefficients $c_i=O(1)$, while the terms at $(k+1)$-st order should be $O(\epsilon)$. This implies that roughly $\lambda^k/\Delta^{k-1}=O(1)$ and $\lambda^{k+1}/\Delta^k=O(\epsilon)$, which implies that $\Delta=O(\epsilon^{-k})$ and $\lambda=O(\epsilon^{-(k-1)})$. Hence $\|T_r\|_2=O(\lambda^r/\Delta^{r-1})=O(\epsilon^{r-k})$. Assuming that the locality of target Hamiltonian $k$ is fixed, as $r$ increases the norm of $\|T_r\|_2$ quickly becomes small enough to justify using the crude bound in Equation \ref{eq:Tr} on the remaining terms of the self-energy expansion. Therefore it suffices to consider $r$ up to $k+d$ for some fixed $d$ such that the total magnitude of the remaining sum is $O(\epsilon^d)$. For $r>k+d$, the upper bound becomes sufficiently small (assuming $\|T_r\|_\infty\rightarrow 0$ as $r\rightarrow\infty$), we use Equation \ref{eq:Tr} to bound the terms from $r=k+d+1$ to infinity.

\section{Efficiently computed tight error bound}\label{sec:errorbound}

In this section we present the details of our techniques. Section \ref{subsec:reduce} introduces the notions that we use for reducing the amount of computation needed for the upper bound. These notions (such as ``configuration" and ``reduced configuration" as will be discussed in Section \ref{subsec:reduce}) are essentially simplified representations of $H$ eigenstates. In Appendix \ref{subsec:example} we provide an explicit example for calculating the norm of $T_2$ that illustrates the uses of these notions, hoping that the presentation be as instructive to the reader as possible. In Section \ref{subsec:algo} we present general algorithms for computing an upper bound to $\|T_r\|_2$ for any fixed $r$. 

In Section \ref{subsec:sharp} we prove that in the case where the terms $H_\text{targ,$i$}$ pairwise commute, our algorithms in fact computes the \emph{exact} value of $\|T_r\|_2$. We accomplish this by first developing further properties of the notions that are introduced in Section \ref{subsec:reduce}, and then show how the algorithms introduced in Section \ref{subsec:algo} use these properties to effectively collect all the terms that contribute to $\|T_r\|$.

\subsection{Reducing the space of summation to polynomial size}\label{subsec:reduce}

For simplicity from here on we let $\lambda_{i,j}=c_i^{1/k}\equiv\lambda_i$ in Equation \ref{eq:Vi} and define the vector $\boldsymbol\lambda=(\lambda_1,\lambda_2,\cdots,\lambda_m)$. Let $E^{(i)}(j)=\frac{j(k-j)}{k-1}\Delta\equiv E_j$ be the energy level of $H^{(i)}$ in Equation \ref{eq:Hi} with $j$ out of $k$ ancillas in $|1\rangle$ state. For a $k$-local target Hamiltonian of $m$ terms the computational basis states of the $km$ ancilla qubits can be represented by $km$-bit strings of the form $s_1s_2\cdots s_m$ with each $s_i$ being a $k$-bit string describing the state of the ancilla qubits in the $i$-th register. For any $\phi\in\{0,1\}^{km}$, we use the notation $E(\phi)=\langle\phi|H|\phi\rangle$ to represent the energy of the ancilla state $|\phi\rangle$. Recall that $H=\sum_{i=1}^mH^{(i)}$ and the $H^{(i)}$ terms pairwise commute. As an explicit connection between the $E(\phi)$ and $E^{(i)}(j)$ notations, defining $h(s)$ as the Hamming weight of a string $s$, we have 
\begin{equation}\label{eq:Ephi}
E(\phi)=\sum_{i=1}^mE^{(i)}(h(s_i)).
\end{equation}

In order to gain more insights about the structure of the $r$-th order term $T_r$ in Equation \ref{eq:T_r}, we insert resolutions of identity $I=\sum_{\phi\in\{0,1\}^{km}}|\phi\rangle\langle\phi|$ between the $V$ and $G$ operators in $T_r$. From Equation \ref{eq:T_r} we get
\begin{equation}\label{eq:Tr_walk}
\begin{array}{ccl}
T_r & = & \displaystyle \sum_{\phi_0,\phi_1,\cdots,\phi_r\in\{0,1\}^{km}}\left(\langle\phi_0|V|\phi_1\rangle\frac{1}{z-E({\phi_1})}\langle\phi_1|V|\phi_2\rangle\cdots\right. \\[0.2in]
& & \displaystyle \makebox[1.7in]{}\left.\cdots\frac{1}{z-E({\phi_{r-1})}}\langle\phi_{r-1}|V|\phi_r\rangle\right)\otimes|\phi_0\rangle\langle\phi_r|. \\[0.2in]
\end{array}
\end{equation}
Note that in the basis of $|\phi\rangle$ states, each matrix block $\langle \phi_0|T_r|\phi_r\rangle$ is a sum of $2^{kmr}$ terms, which is an enormous amount of computation. However, we note that a majority of summants are in fact zero.
In order for a sequence of states $(\phi_0,\phi_1,\cdots,\phi_r)$ to have non-zero contribution to the sum in Equation \ref{eq:Tr_walk}, there are certain conditions that must be satisfied:
\begin{enumerate}
\item Because the self-energy $\Sigma_-(z)$ is restricted to the low energy subspace $\mathcal{L}_-$ of the unperturbed Hamiltonian $H$, both $|\phi_0\rangle$ and $|\phi_r\rangle$ must be restricted to $\mathcal{L}_-=\bigotimes_{i=1}^m\mathcal{L}_{-.i}=\text{span}\{|0^k\rangle,|1^k\rangle\}^{\otimes m}$. In other words, they must be of the form $s_1s_2\cdots s_m$ with each $s_i\in\{0^k,1^k\}$;\label{cond:Lm}
\item Because each $V^{(i)}$ contains only single Pauli $X$ operators acting on the ancilla qubits, for any pair of ancilla states $|\phi_i\rangle$, $|\phi_j\rangle$, the matrix block $\langle\phi_i|V|\phi_j\rangle$ is nonzero iff $\phi_i$ and $\phi_j$ differ by one and only one bit;\label{cond:diff}
\item Because the $V_+$ and $G_+$ terms in $T_r$ are projections onto the high energy subspace $\mathcal{L}_+$ of $H$, all the intermediate states $\phi_1$ through $\phi_{r-1}$ must also belong to $\mathcal{L}_+$. In other words, they must \emph{not} be of the form $s_1s_2\cdots s_m$ with each $s_i\in\{0^k,1^k\}$.\label{cond:Lp}
\end{enumerate}
Conditions \ref{cond:Lm} and \ref{cond:Lp} are reminiscent of Goldstone's theorem in quantum many-body physics \cite{FW71,Goldstone267}, where all non-zero contributions to the spectral difference between the perturbed and the unperturbed systems is a summation of ``connected diagrams" \emph{i.e.\ }the state with no particles or  holes present can never occur as an intermediate state because the resulting matrix element will contain disconnected parts.
For condition \ref{cond:diff} above if $\phi_i$ and $\phi_j$ differ at bit which belongs to the $p$-th register of $k$ ancilla qubits, then 
\begin{equation}\label{eq:lp}
\|\langle\phi_i|V|\phi_j\rangle\|_\infty=\lambda_p.
\end{equation} 
Let $\mathcal{W}_r$ be the set of sequences $(\phi_0,\phi_1,\cdots,\phi_r)$ that satisfy the above three conditions. Then the summation in Equation \ref{eq:Tr_walk} can be replaced by a summation over $\mathcal{W}_r$. The following Lemma states that to find an upper bound to $\|T_r\|_2$ from Equation \ref{eq:Tr_walk} it suffices to consider a subset of sequences with \emph{fixed} $\phi_0=0^{km}$, which is a string of $km$ zeros. For convenience from here on we use ``$\phi\in\mathcal{L}_-$" to as a shorthand for ``$\phi\in\{0,1\}^{mk}:|\phi\rangle\in\mathcal{L}_-$".
$\quad$\\
\begin{lemma}\label{lemma:0k}
For the $r$-th order term $T_r$ as written in Equation \ref{eq:T_r}, we have
\begin{equation}\label{eq:Tr_zero_phir}
\begin{array}{c}
\|T_r\|_2 \le \sum_{\phi_r\in\mathcal{L}_-}\|\langle 0^{mk}|T_r|\phi_r\rangle\|_\infty.
\end{array}
\end{equation}
\end{lemma}
\noindent{\bf Proof.}
Let the operator $S_i=\bigotimes_{j=1}^k X_{i,j}$, recalling the subscript notation $``{i,j}"$ means the $j$-th ancilla in the $i$-th register (Equation \ref{eq:Vi}). Then for a state $|\phi\rangle$ of the $mk$ ancillary qubits, $S_i$ acting on $|\phi\rangle$ flips all the bits in the $i$-th register. From Equation \ref{eq:Vi} it is clear that for any sequence of $S_i$ operations 
\begin{equation}\label{eq:S_il}
S=S_{i_1}S_{i_2}\cdots S_{i_\ell}
\end{equation}
with $i_1,i_2,\cdots,i_\ell\in[m]$, we have $SVS =V$. Additionally, the set of all ancilla states $|\phi\rangle\in\mathcal{L}_-$ with $\phi\in\{0,1\}^{mk}$ is invariant with respect to $S_i$ for any $i$. Finally, substituting $V$ with $SVS$ in Equation \ref{eq:Tr_walk} leads to 
\begin{equation}\label{eq:STS_phi}
\langle\phi_0|ST_rS|\phi_r\rangle=\langle\phi_0|T_r|\phi_r\rangle. 
\end{equation}
Because we could express any $\phi\in\mathcal{L}_-$ as $|\phi\rangle=S|0^{mk}\rangle$ for some $S$ with the form in Equation \ref{eq:S_il}, we could further write $T_r$ as
\begin{equation}\label{eq:Tr_reorg}
\begin{array}{ccl}
T_r & = &\displaystyle \sum_{\phi_r\in\mathcal{L}_-}\sum_S\langle 0^{km}|ST_r|\phi_r\rangle S|0^{km}\rangle\langle \phi_r| \\
& = & \displaystyle \sum_{\phi_r\in\mathcal{L}_-}\sum_S\langle 0^{km}|T_rS|\phi_r\rangle S|0^{km}\rangle\langle \phi_r| \\
& = & \displaystyle \sum_{\phi_r\in\mathcal{L}_-}\langle 0^{km}|T_r|\phi_r\rangle\otimes
\underbrace{\left(\sum_S S|0^{km}\rangle\langle\phi_r|S\right)}_{(*)}.
\end{array}
\end{equation}
Here the summation $\sum_S$ is over all operators of the form in Equation \ref{eq:S_il}. Going from the first line to the second we have used Equation \ref{eq:STS_phi}. Going from the second line to the third is a substitution of variable $|\phi_r\rangle\rightarrow S|\phi_r\rangle$. Note that the term $(*)$ in Equation \ref{eq:Tr_reorg} is a sum of projectors that are orthogonal to each other (\emph{i.e.\ }each pair multiply to zero), implying that the norm of $(*)$ is always one. Therefore to find an upper bound to $\|T_r\|_2$ from Equation \ref{eq:Tr_walk} it suffices to consider a \emph{fixed} $\phi_0=0^{km}$, which is a string of $km$ zeros (For convenience from here on we use ``$\phi\in\mathcal{L}_-$" to as a shorthand for ``$\phi\in\{0,1\}^{mk}:|\phi\rangle\in\mathcal{L}_-$"):
\begin{equation}\label{eq:Tr_zero_phir}
\begin{array}{c}
\|T_r\|_2 \le \displaystyle \|T_r\|_\infty\le\max_{\phi_0\in\mathcal{L}_-} \sum_{\phi_r\in\mathcal{L}_-}\|\langle \phi_0|T_r|\phi_r\rangle\|_\infty = \sum_{\phi_r\in\mathcal{L}_-}\|\langle 0^{mk}|T_r|\phi_r\rangle\|_\infty.
\end{array}
\end{equation}
Here in Equation \ref{eq:Tr_zero_phir} the first $\le$ uses $\|T_r\|_\infty=\|T_r\|_1$ implied by the hermiticity of $T_r$ and the property $\|T_r\|_2^2\le\|T_r\|_1\cdot\|T_r\|_\infty$, where $\|\cdot\|_1$ is the 1-norm of a matrix defined as ``maximum column sum" $\|A\|_1=\max_j\sum_i|\langle i|A|j\rangle|$. The second $\le$ uses the definition of the $\infty$-norm. The final equality in Equation \ref{eq:Tr_zero_phir} comes from the invariance property described in Equation \ref{eq:Tr_reorg}. \hfill{$\square$}

\begin{figure}
\begin{center}
\includegraphics[scale=0.95]{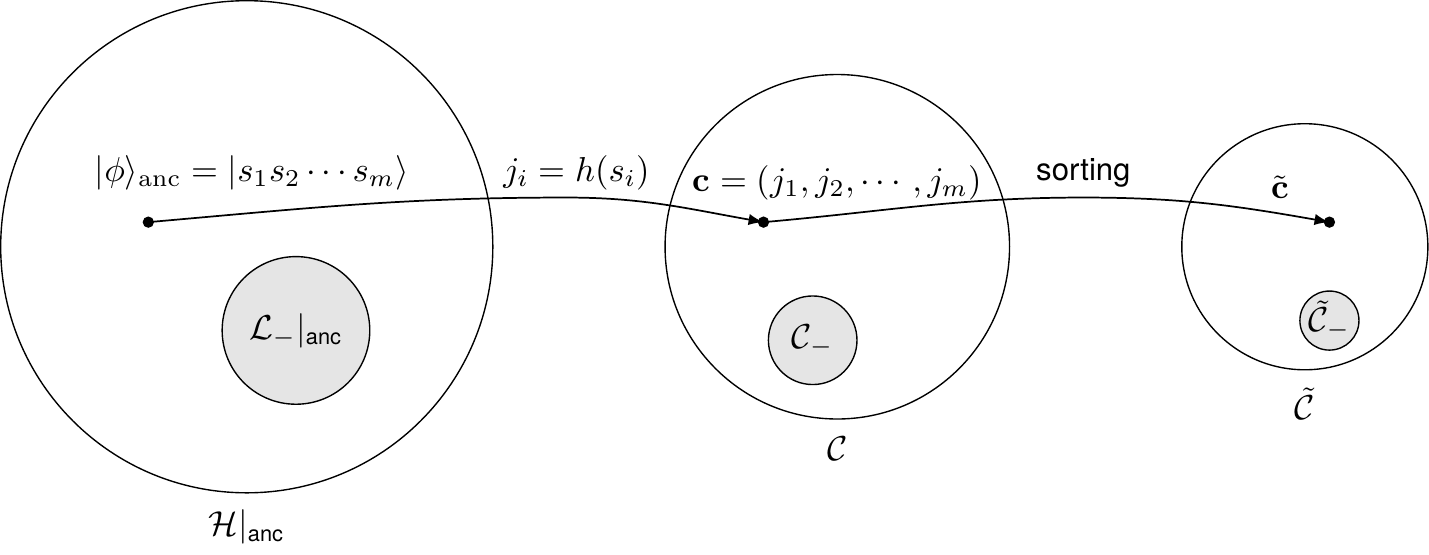}
\fcaption{Relationship between elements of the various spaces that are relevant to our discussion in Section \ref{subsec:reduce}. We use the notation $|_\text{anc}$ to represent the restriction of the Hibert spaces on which the gadget Hamiltonian acts to the ancilla registers. Here the $s_i$'s are $k$-bit strings, and $h(\cdot)$ is the Hamming weight of a string. We highlight the subspaces (or subsets) that correspond to the low energy subspace of the unperturbed Hamiltonian $H$. $\mathcal{C}_-$ consists of all the configurations of ancilla states in $\mathcal{L}_-|_\text{anc}$ and $\tilde{\mathcal{C}}_-$ consists of the reduced configurations of those in $\mathcal{C}_-$.}
\label{fig:reduce}
\end{center}
\end{figure}

$\quad$\\
We further partition $\mathcal{W}_r$ into subsets according to different combinations of $\phi_0$ and $\phi_r$. Denote $W_r(\phi_0,\phi_r)$ as the subset of sequences in $\mathcal{W}_r$ that starts from $\phi_0$ and ends at $\phi_r$. For a given sequence $(\phi_0,\phi_1,\cdots,\phi_r)$, let $p_i\in[m]$ be the index of the ancilla register that contains the bit where $\phi_i$ differs from $\phi_{i-1}$. Using Equation \ref{eq:lp}, the norm of each matrix block $\langle\phi_0|T_r|\phi_r\rangle$ can be bounded from above by
\begin{equation}\label{eq:t0r_bound}
\|\langle\phi_0|T_r|\phi_r\rangle\|_\infty\le\sum_{\mathcal{W}_r(\phi_0,\phi_r)}
\underbrace{
\lambda_{p_1}\cdot\frac{1}{|z-E(\phi_1)|}\cdot\lambda_{p_2}\cdots\lambda_{p_{r-1}}\cdot\frac{1}{|z-E(\phi_{r-1})|}\cdot\lambda_{p_r}
}_{\equiv t_\phi(\phi_0,\cdots,\phi_r)}
\end{equation}
where the weight function $t_\phi$ describes the contribution, or the ``weight" of a specific sequence in the sum.
Regardless of the restriction to $\mathcal{W}_r$, evaluating the upper bound in Equation \ref{eq:t0r_bound} with brute-force enumeration of all possible intermediate steps $\phi_1$, $\cdots$, $\phi_{r-1}$ would still lead to a computational cost that is exponential in the number of registers $m$. However, by exploiting the structure of the gadget Hamiltonian we could reduce it to poly$(m)$ for any fixed order $r$ of perturbation theory. Such reduction is accomplished by introducing a sequence of two mappings ${\bf c}$ and $\tilde{\bf c}$ (Figure \ref{fig:reduce}) which we will introduce in the following discussion.

\begin{definition}[Configuration]\label{def:config}
For a state $|\phi\rangle$ with $\phi=s_1s_2\cdots s_m$ where each $s_i$ describes the state of a $k$-qubit register, we define the vector ${\bf c}(\phi)=(j_1,j_2,\cdots,j_m)$ with $j_i=h(s_i)$ as the \emph{configuration} of a state $|\phi\rangle$. Each element $j_i$ of the configuration corresponds to energy level $E^{(i)}(j_i)$ of the term $H^{(i)}$. 
\end{definition}

Previously we have defined the set $\mathcal{W}_r$ as the set of $r$-step sequences $(\phi_0,\phi_1,\cdots,\phi_r)$ that contribute non-trivially to $T_r$. Since each bit string $\phi$ is associated with a configuration ${\bf c}$, each sequence of ancilla states (bit strings) is naturally associated with a sequence of configurations $({\bf c}_0,{\bf c}_1,\cdots,{\bf c}_r)$. Similar to $\mathcal{W}_r$, we define $\mathcal{W}_r^{\bf c}$ as the set of $r$-step sequences $({\bf c}_0,{\bf c}_1,\cdots,{\bf c}_r)$ that correspond to the $r$-step sequences in $\mathcal{W}_r$. We then let $\mathcal{W}_r^{\bf c}({\bf c}_0,{\bf c}_r)$ be the set of $r$-step configuration sequences that starts from specific values of ${\bf c}_0$ and ${\bf c}_r$.  With the Definition \ref{def:config} we could rewrite the sum in Equation \ref{eq:t0r_bound} as
\begin{equation}\label{eq:Tr_c0cr}
\|\langle\phi_0|T_r|\phi_r\rangle\|_\infty\le\sum_{\mathcal{W}_r^{\bf c}({\bf c}_0,{\bf c}_r)}
\sum_{{\bf c}(\phi_i)={\bf c}_i}t_\phi(\phi_0,\cdots,\phi_r)
=\sum_{\mathcal{W}_r^{\bf c}({\bf c}_0,{\bf c}_r)}t_{\bf c}({\bf c}_0,\cdots,{\bf c}_r)
\end{equation}
where $t_{\bf c}({\bf c}_0,\cdots,{\bf c}_r)$ is the total weight of all sequences $(\phi_0,\cdots,\phi_r)$ for which ${\bf c}(\phi_i)={\bf c}_i$. 

Note that unlike Equation \ref{eq:t0r_bound} which sums over states that dwell in a space of dimension $2^{mk}$, the summation in Equation \ref{eq:Tr_c0cr} sums over configurations, which dwells in a space of dimension $O(k^m)$. This is an exponential reduction with respect to $k$, but the dimension is still exponential in $m$ nonetheless. To reduce the dimension further, we first note that the energy of a configuration is invariant with respect to permutations of the ancilla registers. We would like to exploit this permutation invariance by restricting to a class of configuration vectors that are \emph{sorted}. Specifically, we put forward the following definition.
\begin{definition}[Reduced configuration]\label{def:red_config}
For a configuration ${\bf c}$ as in Definition \ref{def:config}, we define the \emph{reduced configuration} of ${\bf c}$, denoted as $\tilde{\bf c}({\bf c})$, as the vector obtained by sorting a configuration ${\bf c}$ in non-decreasing order. 
\end{definition}
Then by definition any set of configurations that differ by only permutations of the ancilla registers share the same reduced configurations. With an elementary inductive argument we can show that while there are $O(k^m)$ configurations, there are only $O(m^k)$ reduced configurations \cite{redsignlemma}. 

Since each sequence of configurations  $({\bf c}_0,{\bf c}_1,\cdots,{\bf c}_r)$ has a corresponding sequence of reduced configurations  $(\tilde{\bf c}_0,\tilde{\bf c}_1,\cdots,\tilde{\bf c}_r)$, we define $\widetilde{\mathcal{W}}_r^{\bf c}$ as the set of reduced configuration sequences derived from all sequences of configurations in $\mathcal{W}_r^{\bf c}$. We then let $\widetilde{\mathcal{W}}_r^{\bf c}(\tilde{\bf c}_0,\tilde{\bf c}_r)$ be the set of $r$-step configuration sequences that starts from specific values of $\tilde{\bf c}_0$ and $\tilde{\bf c}_r$. The sum in Equation \ref{eq:Tr_c0cr} could then be further rewritten as
\begin{equation}\label{eq:Tr_red_c0cr}
\|\langle\phi_0|T_r|\phi_r\rangle\|_\infty\le \sum_{\widetilde{W}_r^{\bf c}(\tilde{\bf c}_0,\tilde{\bf c}_r)}
\sum_{\tilde{\bf c}({\bf c}_i)=\tilde{\bf c}_i}t_{\bf c}({\bf c}_0,\cdots,{\bf c}_r)
=\sum_{\mathcal{W}_r^{\bf c}({\bf c}_0,{\bf c}_r)}\tilde{t}_{\bf c}(\tilde{\bf c}_0,\cdots,\tilde{\bf c}_r)
\end{equation}
where $\tilde{t}_{\bf c}(\tilde{\bf c}_0,\cdots,\tilde{\bf c}_r)$ sums over all sequences $({\bf c}_0,{\bf c}_1,\cdots,{\bf c}_r)$ for which $\tilde{\bf c}({\bf c}_i)=\tilde{\bf c}_i$. 
Figure \ref{fig:reduce} summarizes the relationship between the space of $H$ eigenstates (restricted to the ancilla qubits), configurations and reduced configurations.

Let $\widetilde{\mathcal{C}}_-$ be the set of reduced configurations corresponding to basis states in $\mathcal{L_-}$. Then since for each ancilla register the low energy subspace is spanned by states with 0 or $k$ qubits in $|1\rangle$ state, we have
\begin{equation}\label{eq:tildec_minus}
\widetilde{\mathcal{C}}_-=\{i\in[m]|(\underbrace{k,k,\cdots,k}_\text{$i$ times},0,0,\cdots,0)\}.
\end{equation}
This property will become useful in our later discussion of Algorithm \ref{alg:perturbbound} in Section \ref{subsec:algo}. Combining Lemma \ref{lemma:0k} and Equation \ref{eq:Tr_red_c0cr} yields an upper bound to $\|T_r\|_2$ with \emph{fixed} $\tilde{\bf c}_0=(0,0,\cdots,0)$ as
\begin{equation}\label{eq:Tr_bound}
\|T_r\|_2\le\sum_{\tilde{\bf c}_r\in\widetilde{\mathcal{C}}_-}\sum_{\widetilde{\mathcal{W}}_r^{\bf c}(\tilde{\bf c}_0,\tilde{\bf c}_r)}\tilde{t}_{\bf c}(\tilde{\bf c}_0,\tilde{\bf c}_1,\cdots,\tilde{\bf c}_{r})
\end{equation}
with the function $t$ defined in Equation \ref{eq:Tr_red_c0cr}.
Equation \ref{eq:Tr_bound} is the basis of our main algorithms for finding an upper bound to $\|T_r\|_2$ to be presented in Section \ref{subsec:algo}. In order to illustrate the definitions and their properties introduced in this section, in Appendix \ref{subsec:example} we present a detailed example showing the procedure for estimating $\|T_2\|_2$. The algorithms for general $\|T_r\|_2$ are discusssed in the subsequent Section \ref{subsec:algo}.

\subsection{Algorithm for computing the upper bound}\label{subsec:algo}

An upper bound to the right hand side of Equation \ref{eq:Tr_bound} can be efficiently evaluated and expressed using \emph{monomial symmetric polynomials} in the $\lambda_i$ coefficients \cite{CK16}. A monomial symmetric polynomial $m_{\bf b}(x_1,x_2,\cdots,x_n)=\sum_{\pi}x_{\pi(1)}^{b_1}x_{\pi(2)}^{b_2}\cdots x_{\pi(k)}^{b_k}$ where ${\bf b}\in\mathbb{N}^k$ is the \emph{partition} of the symmetric polynomial and $\pi:[n]\mapsto[k]$ is an arrangement of $k$ elements among $n$ elements. In order to address the combinatorics involved in summing over $\tilde{\bf c}$ sequences, we also need a $(k+1)\times(k+1)$ matrix $M$ with element $M_{ij}$ being the number of possible ways to cause a transition from an $H$ eigenstate with energy $E_i$ to an eigenstate with energy $E_j$ by one application of the perturbation $V$. For the unperturbed Hamiltonian defined according to Equation \ref{eq:Hi}, we have
\begin{equation}\label{eq:M}
M_{ij}= \left\{
\begin{array}{lc}
i & \text{if $j=i-1$} \\
k-i & \text{if $j=i+1$} \\
0 & \text{otherwise}.
\end{array}\right.
\end{equation}

With the above definition in place, here we present a simple method for evaluating the sum in Equation \ref{eq:Tr_red_c0cr}. Consider a fixed sequence of reduced configurations $\tilde{\bf c}_0$, $\tilde{\bf c}_1$, $\cdots$, $\tilde{\bf c}_r$ with $\tilde{\bf c}_0=({0,0,\cdots,0})$. We compute $w=\tilde{t}_{\bf c}(\tilde{\bf c}_1,\tilde{\bf c}_2,\cdots,\tilde{\bf c}_{r})$, the term in Equation \ref{eq:Tr_red_c0cr} corresponding to $\tilde{\bf c}_0$, $\tilde{\bf c}_1$, $\cdots$, $\tilde{\bf c}_r$, by Algorithm \ref{alg:walkbound}, which can be considered as a procedure for computing the symmetric polynomial that is an upper bound to the right hand side of \eqref{eq:2nd_walkbound} for general $r$. The discussion in Appendix \ref{subsec:example} may be used as an example for bounding the second order term\footnote{
From \eqref{eq:seq} in Appendix \ref{subsec:example} we can see that $M_{01}=3$. The sequences in \eqref{eq:seq} give rise to the term whose norm is $3(\lambda_1^2+\lambda_2^2)\cdot\frac{1}{z-E_1}$, which can be interpreted as a symmetric polynomial (recall that we have fixed $\tilde{\bf c}_0=(0,0)$)
\begin{equation}\label{eq:2nd_walkbound}
\frac{1}{|z-E_1|}\cdot\Omega_1\Omega_2m_{(2)}(\lambda_1,\lambda_2)=\sum_{\substack{\text{$\tilde{\bf c}_1$, $\tilde{\bf c}_2$ such that} \\ (\tilde{\bf c}_0,\tilde{\bf c}_1,\tilde{\bf c}_2)\in\widetilde{W}_2^{\bf c}}}t(\tilde{\bf c}_0,\tilde{\bf c}_1,\tilde{\bf c}_2)
\end{equation}
with $\Omega_1=M_{01}=3$ and $\Omega_2=M_{10}=1$. Note that the right hand side of \eqref{eq:2nd_walkbound} matches with that of Equation \ref{eq:Tr_bound} for $r=2$.
}.

\begin{algorithm}
\begin{enumerate}
\item Let $\tilde{\bf c}_0=(\underbrace{0,0,\cdots,0})$. \label{step:init}
\item Check if the following holds: \label{step:check}
\begin{enumerate}
\item $\tilde{\bf c}_r\notin\widetilde{\mathcal{C}}_-$; \label{step:check_cr}
\item there is any pair of reduced configurations $\tilde{\bf c}_i=\tilde{\bf c}_{i+1}$. \label{step:check_equal}
\item any of the reduced configuration $\tilde{\bf c}_i$ with $i\in[r-1]$ satisfies ${\bf n}(\tilde{\bf c}_i)\notin\mathcal{N}_+$; \label{step:check_plus}
\end{enumerate}
If either of the conditions hold, $w=0$ and return.
\item For each $i=0,\cdots,r$, introduce a partition vector ${\bf b}_i$ of length at most $i$ and a mapping $\mu_i:\tilde{\bf c}_i\mapsto{\bf b}_i$ that maps some elements of $\tilde{\bf c}_i$ to ${\bf b}_i$. We would like each element in ${\bf b}_i$ to have a unique pre-image in ${\bf c}_i$, because intuitively, the $q$-th element of ${\bf b}_i$, denoted as $b_{i,q}$, represents how many times the register $\mu^{-1}(b_{i,q})$ has been acted on by $V$ during the sequence. \label{step:ds}
\item Start from ${\bf b}_0=\emptyset$ and $\mu_0=\emptyset$. We scan from $\tilde{\bf c}_1$ through $\tilde{\bf c}_r$ and update the $\mu$ and ${\bf b}$ assignments in the following way. Suppose we have already computed $\mu_i$ and ${\bf b}_i$. Then we find the element $\tilde c_{i+1,s}$ in $\tilde{\bf c}_{i+1}$ that differs from the corresponding element $\tilde c_{i,s}$ in $\tilde{\bf c}_i$. \label{step:scan}
\begin{enumerate}
\item If $\tilde c_{i,s}$ is not in the domain of $\mu_i$ (implying $\tilde{c}_{i,s}=0$ since the $s$-th element of $\tilde{\bf c}$ has not been modified by the algorithm before), let ${\bf b}_{i+1}={\bf b}_i\cup\{b_t\}$ with $b_t=1$, $\tilde c_{i+1,s}=1$ and $\mu_{i+1}=\mu_i\cup\{\tilde c_{i+1,s}\mapsto b_t\}$; \label{step:notin}
\item If $\tilde c_{i,s}$ is in the domain of $\mu_i$, then first let ${\bf b}_{i+1}={\bf b}_i$ and $\mu_{i+1}=\mu_i$ and then increment $\mu_{i+1}(\tilde{\bf c}_{i+1})$ by 1; \label{step:in}
\item Compute and store $\Omega_i=M_{xy}$ with $x=\tilde c_{i,s}$ and $y=\tilde c_{i+1,s}$. \label{step:omega}
\end{enumerate}
\item Return $w(\tilde{\bf c}_1,\tilde{\bf c}_2,\cdots,\tilde{\bf c}_{r-1})=\left(\Pi_{i=1}^{r-1}{|z-E(\tilde{\bf c}_i)|}^{-1}\right)(\Pi_{i=1}^r\Omega_i)m_{{\bf b}_r}(\lambda_1,\lambda_2,\cdots,\lambda_m)$.
\end{enumerate}
\caption{\textsc{: $w=$WalkBound($\tilde{\bf c}_1$, $\tilde{\bf c}_2$, $\cdots$, $\tilde{\bf c}_r$)}}
\label{alg:walkbound}
\end{algorithm}

In Algorithm \ref{alg:walkbound}, we assume (step \ref{step:init}) that $\tilde{\bf c}_0$ is the all-zero vector, as a consequence of the discussion that leads to Equation \ref{eq:Tr_zero_phir}. Then in step \ref{step:check} we check if the input sequence of reduced configurations would actually produce a non-zero contribution to $T_r$ by examining the three criteria listed in Section \ref{subsec:reduce} after Equation \ref{eq:Tr_walk}. Step \ref{step:check_cr}, \ref{step:check_equal} and \ref{step:check_plus} examines violation of each of the three criteria respectively. If any of the criteria is violated, return $w=0$ since the input sequence does not contribute non-trivially to $T_r$. Step \ref{step:ds} introduces the data structure used for representing a monomial symmetric polynomial, which includes 1) a partition vector $\bf b$, 2) a reduced configuration $\tilde{\bf c}$ as given by the input sequence and 3) an injective partial function $\mu:\tilde{\bf c}\mapsto{\bf b}$. The partition $\bf b$ is needed for computing the symmetric polynomial $m_{\bf b}(\boldsymbol\lambda)$, while $\tilde{\bf c}$ and $\mu$ are needed for guiding the computation of $\bf b$. As shown in step \ref{step:scan}, depending on how the reduced configuration changes from the current $i$-th step to the $(i+1)$-st, the data structure for the new step is updated. Because at this point of the algorithm $\tilde{\bf c}_i$ and $\tilde{\bf c}_{i+1}$ differ by exactly one element (step \ref{step:check} has excluded all invalid sequences of reduced configurations), what matters is then the position of the differing element, which we call $\tilde{c}_{i,s}$. There are two possibilities, each handled by step \ref{step:notin} and \ref{step:in}. Recall that the value of each element in a reduced configuration stands for the energy level of an ancilla register. Making a transition from energy level $x=\tilde{c}_{i,s}$ to $y=\tilde{c}_{i+1,s}$ induces a combinatorial factor $M_{xy}$ which is handled in step \ref{step:omega}. 

The steps for computing $\mu_r$ and ${\bf b}_r$ takes $O(r)$ time. In the final step, evaluating the symmetric polynomial in $w$ takes $O(m^r)$ time. Hence for a fixed $r$, the total time needed for computing the upper bound $w$ for a \emph{fixed} sequence of reduced configurations is polynomial in $m$. We outline the algorithm for computing the tight error bound in Algorithm \ref{alg:perturbbound}. Recall the definition of $\boldsymbol\lambda$ at the beginning of Section \ref{subsec:reduce}. 

\begin{algorithm}
\begin{enumerate}
\item If $r\ge k$, compute integers $p=\lfloor r/k\rfloor$ and $q=r-p$. Otherwise let $p=0$, $q=0$. \label{step:pq}
\item If $q$ is odd, let $\gamma_r=0$. Otherwise, compute 
\begin{equation}
\displaystyle\gamma_r=\sum_{\substack{\text{$\tilde{\bf c}_1$, $\cdots$, $\tilde{\bf c}_{r-1}$ such that} \\ (\tilde{\bf c}_0,\cdots,\tilde{\bf c}_{r-1},\tilde{\bf c}_0)\in\widetilde{\mathcal{W}}_r^{\bf c}}}\textsc{WalkBound($\tilde{\bf c}_1$, $\cdots$, $\tilde{\bf c}_{r-1}$, $\tilde{\bf c}_0$)}.
\end{equation} \label{step:alpha_r}
\item If $r\ge k$, then for each $i$ from 1 to $\min\{p,m\}$ \label{step:outer_loop}
\begin{enumerate}
\item Let $\tilde{\bf c}_r=(\underbrace{k,k,\cdots,k}_\text{$i$ times}, \underbrace{0,0,\cdots,0}_\text{$m-i$ times})$;
\item Compute $\displaystyle\gamma_{i,r}=\sum_{\substack{\text{$\tilde{\bf c}_1$, $\cdots$, $\tilde{\bf c}_{r-1}$ such that} \\ (\tilde{\bf c}_0,\cdots,\tilde{\bf c}_{r-1},\tilde{\bf c}_r)\in\widetilde{\mathcal{W}}_r^{\bf c}}}$\textsc{WalkBound($\tilde{\bf c}_1$, $\cdots$, $\tilde{\bf c}_{r-1}$, $\tilde{\bf c}_r$)}.\label{step:gamma}
\end{enumerate}
Otherwise let $\gamma_{i,r}=0$ for all $i$.
\item Return $\tau_r=\gamma_r+\sum_{i=1}^{\min\{p,m\}}\gamma_{i,r}$.
\end{enumerate}
\caption{\textsc{: $\tau_r=$PerturbBound($r$,$k$,$\boldsymbol\lambda$,$M$)}}
\label{alg:perturbbound}
\end{algorithm}

Algorithm \ref{alg:perturbbound} essentially computes the right hand side of Equation \ref{eq:Tr_bound}. Step \ref{step:pq} computes parameters $p$ and $q$ that are relevant to the the general structure of $\mathcal{W}_r$ introduced in the beginning of Section \ref{subsec:reduce}. Recall the three criteria for sequences $(\phi_0,\phi_1,\cdots,\phi_r)$ in $\mathcal{W}_r$: $\phi_0$ and $\phi_r$ must be of the form $s_1s_2\cdots s_m$ with each $s_i\in\{0^k,1^k\}$ while $\phi_1$, $\cdots$, $\phi_{r-1}$ must not be of this form and $\phi_i$, $\phi_{i+1}$ must differ by one and only one bit for $i=0,\cdots,r-1$. For $r<k$, clearly $\mathcal{W}_r$ could only contain sequences where $\phi_0=\phi_r$, resulting in $T_r$ being proportional to the identity operator in $\mathcal{L}_-$ with the proportional constant being $\gamma_r$ (Equation \ref{eq:sigmaz_heff}). For $r\ge k$, it is then possible that $\phi_r=s_1s_2\cdots s_m$ with at most $p=\lfloor r/k\rfloor$ substrings $s_i=1^k$. Therefore one needs to also take sequences with these $\phi_r$ possibilities into account. Step \ref{step:alpha_r} computes the magnitude of the term in $T_r$ that is proportional to identity in the low energy subspace $\mathcal{L}_-$, by summing over all sequences of reduced configurations that starts from $\tilde{\bf c}_0$ and ends at $\tilde{\bf c}_0$. For $r\le k$, the $\gamma_r$ computed here is precisely the leading coeffcients $\gamma_r$ in Equation \ref{eq:sigmaz_heff}. Looping over all viable sequences $(\tilde{\bf c}_0,\tilde{\bf c}_1,\cdots,\tilde{\bf c}_{r-1},\tilde{\bf c}_0)$ induces $O(m^{k(r-1)})$ in the computational cost. Step \ref{step:outer_loop} completes the remainder of the outer summation in Equation \ref{eq:Tr_bound} over the space $\widetilde{\mathcal{C}}_-$ for $\tilde{\bf c}_r$. In the case where $r<k$ this step is entirely skipped and there will be no $\gamma_{i,r}$ values computed. There are in total at most $m$ iterations in the step and each iteration sum over at most $O(m^{kr})$ sequences $(\tilde{\bf c}_0,\tilde{\bf c}_1,\cdots,\tilde{\bf c}_r)$. Hence the total runtime of Algorithm \ref{alg:perturbbound} scales as $O(m^{kr+1})$, which is polynomial in the number $m$ of ancilla registers. 

\subsection{Sharpness of the upper bound}\label{subsec:sharp}

Algorithm \ref{alg:perturbbound} is useful for computing the exact value of $\gamma_r$ in $T_r$ with $r<k$ (Equation \ref{eq:sigmaz_heff}). Here we show that for the special case where the terms $H_{\text{eff},i}$ pairwise commute, Algorithm \ref{alg:perturbbound} also allows one to efficiently compute the exact value of $\|T_r\|_2$ for any $r\ge k$. This is a stronger claim than the one implied by Equation \ref{eq:Tr_bound} and we state it precisely in the following Theorem \ref{th:sharpness}. 
\begin{theorem}\label{th:sharpness}
Given the $k$-body target Hamiltonian $H_\text{targ}=\sum_{i=1}^mc_iH_\text{targ,$i$}$ as defined in Section \ref{sec:pt}, the gadget Hamiltonian $\tilde{H}=H+V$ as defined in Equations \ref{eq:Hi} and \ref{eq:Vi}, and the self energy expansion $\Sigma_-(z)$ shown in Equation \ref{eq:sigmaz_heff}. If for any $i,j\in\{1,\cdots,m\}$, $[H_\text{targ,$i$},H_\text{targ,$j$}]=0$, then for any $r\ge 2$ and $k\ge 3$, we have
\begin{equation}\label{eq:Tr_equal}
\|T_r\|_2=\textsc{PerturbBound($r$,$k$,$\boldsymbol\lambda$,$M$)}
\end{equation}
where $\boldsymbol\lambda$ is defined in Section \ref{subsec:reduce} and $M$ is defined in Equation \ref{eq:M}.
\end{theorem}

\begin{figure}
\begin{center}
\includegraphics[scale=1]{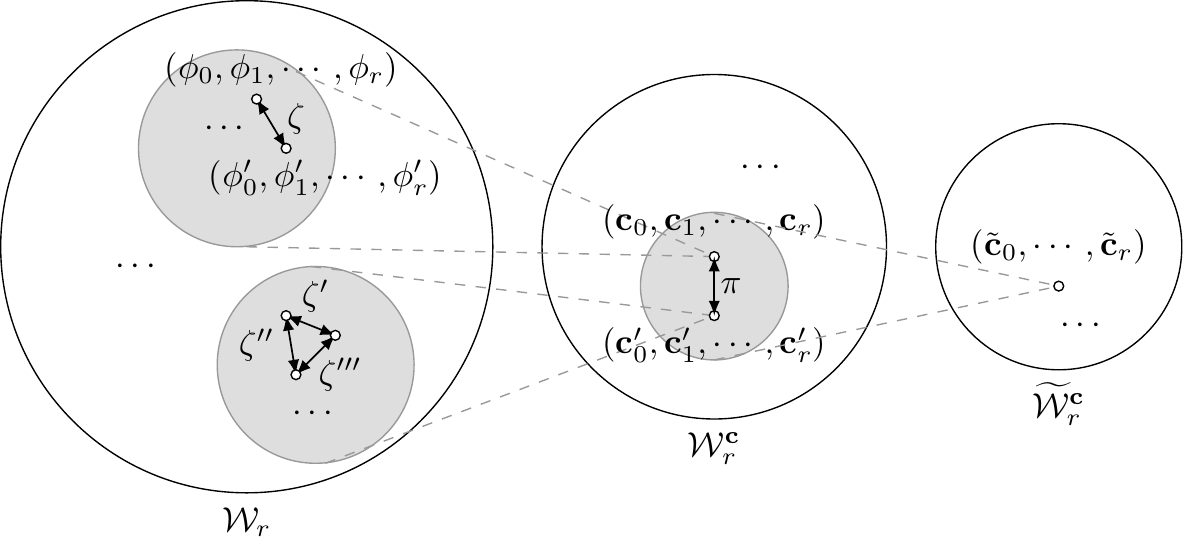}
\fcaption{The hierarchy of equivalence classes that relates the sets $\mathcal{W}_r$, $\mathcal{W}_r^{\bf c}$, and $\widetilde{\mathcal{W}}_r^{\bf c}$. Each light shaded circle represents an equivalence class inside the set. For $\mathcal{W}_r$, each node represents a sequence and sequences in the same equivalence class are related by a ``local permutation'' (defined formally in Section \ref{subsec:sharp}) $\zeta$ that only permutes qubits in the same ancilla register. Each equivalence class in $\mathcal{W}_r$ then maps to a sequence of configurations in $\mathcal{W}_r^{\bf c}$, as shown with the dashed lines, and each sequence of configurations in $\mathcal{W}_r^{\bf c}$ also belongs to an equivalence class where the sequences are related by permutation $\pi$ of the elements of the configuration \emph{i.e.\ }ancilla registers. Each equivalence class in $\mathcal{W}_r^{\bf c}$ then corresponds to an element in $\widetilde{\mathcal{W}}_r^{\bf c}$.}
\label{fig:W_reduce}
\end{center}
\end{figure}

Before proving Theorem \ref{th:sharpness}, we would like to first establish a few properties of the sets $\mathcal{W}_r$, $\mathcal{W}_r^{\bf c}$, and $\widetilde{W}_r^{\bf c}$, which are introduced in Section \ref{subsec:reduce}. Recall that $\mathcal{W}_r$ is a collection of sequences $(\phi_0,\phi_1,\cdots,\phi_r)$ with each $\phi_i$ being a $km$-bit string which we write as $s_1s_2\cdots s_m$, $s_j\in\{0,1\}^k$. Let $\zeta$ be a permutation that only involves permuting bits inside the same $k$-bit substrings $s_j$. In other words, for any bit $b\in\phi=s_1s_2\cdots s_m$, let $j$ be such that $b\in s_j$, then $\zeta(b)\in s_j$ always holds. We call such $\zeta$ a \emph{local permutation}. For any sequence $W=(\phi_0,\phi_1,\cdots,\phi_r)$, we use the notation $\zeta(W)=(\zeta(\phi_0),\zeta(\phi_1),\cdots,\zeta(\phi_r))$ to mean a sequence produced by applying the permutation $\zeta$ onto the string at every intermediate step. We further say that for two sequences $A$ and $B$ in $\mathcal{W}_r$, $A\sim B$ if there is a local permutation $\zeta$ such that $A=\zeta(B)$. Clearly the relation $\sim$ is symmetric, reflexive and transitive. Hence $\sim$ is an equivalence relation that partitions $\mathcal{W}_r$ into equivalence classes (Figure \ref{fig:W_reduce}). Because local permutations do not alter the Hamming weight of any $k$-bit substrings $s_i$, each equivalence class in $\mathcal{W}_r$ corresponds to an element in $\mathcal{W}_r^{\bf c}$ since a configuration ${\bf c}$ is constructed based on Hamming weights of substrings in a state $|\phi\rangle$ (Figure \ref{fig:reduce}). In the set $\mathcal{W}_r^{\bf c}$ we could also define an equivalence relation $\sim^{\bf c}$ based on permutation over the ancilla registers \emph{i.e.\ }the elements of the configuration vector. For $A$, $B$ in $\mathcal{W}_r^{\bf c}$, we say $A\sim^{\bf c} B$ if there is a permutation $\pi$ over ancilla registers such that $A=\pi(B)$. Here the definition of $\pi$ acting on a sequence of configurations is analogous to $\zeta$ on a sequence of ancilla states and we omit the details. Since by definition reduced configurations are obtained from sorting the elements of configurations, each equivalence class in $\mathcal{W}_r^{\bf c}$ naturally corresponds to a sequence of reduced configuration in $\widetilde{\mathcal{W}}_r^{\bf c}$. 

$\quad$\\
\noindent{\bf Proof of Theorem \ref{th:sharpness}.} 
Because we could express any $\phi\in\mathcal{L}_-$ as $|\phi\rangle=S|0^{mk}\rangle$ for some $S$ with the form in Equation \ref{eq:S_il}, we could write $T_r$ as (cf.\ Equation \ref{eq:Tr_reorg})
\begin{equation}
T_r=\sum_{S_1,S_2}\langle 0^{km}|S_1T_rS_2|0^{km}\rangle S_1|0^{km}\rangle\langle 0^{km}|S_2
\end{equation}
where the summation is over any pair of operators $S_1$, $S_2$ of the form in Equation \ref{eq:S_il}. We could further split $T_r$ as a sum of diagonal and off-diagonal components:
\begin{equation}
\begin{array}{ccl}
T_r & = & \displaystyle \sum_S \langle 0^{km}|ST_rS|0^{km}\rangle + \sum_{S_1}\sum_{S_2\neq S_1}\langle 0^{km}|T_rS_1S_2|0^{km}\rangle S_1|0^{km}\rangle\langle 0^{km}|S_2 \\
& = & \displaystyle \sum_S\langle 0^{km}|T_r|0^{km}\rangle S|0^{km}\rangle\langle 0^{km}|S + \sum_{S_1}\sum_{S\neq I}\langle 0^{km}|T_rS|0^{km}\rangle S_1|0^{km}\rangle\langle 0^{km}|S_1S \\
& = & \displaystyle \langle 0^{km}|T_r|0^{km}\rangle \sum_S S|0^{km}\rangle\langle 0^{km}|S + \sum_{S\neq I}\langle 0^{km}|T_rS|0^{km}\rangle\sum_{S_1}S_1(|0^{km}\rangle\langle 0^{km}|S)S_1 \\
& = & \displaystyle \sum_S O_{S,r}\otimes \Pi_S
\end{array}
\end{equation}
where going from the first line to the second we have used Equation \ref{eq:STS_phi} on the first term and applied the substitution $S_2=S_1S$ to the second term. Going from the second line to the third is merely a relocation of the summation so that the form of the expression can be more easily recognized as the last line with 
\begin{equation}
O_{S,r}=\langle 0^{km}|T_rS|0^{km}\rangle\makebox[0.6cm]{}\text{and}\makebox[0.6cm]{}
\Pi_S=\sum_{S'}S'(|0^{km}\rangle\langle 0^{km}|S)S'. 
\end{equation}
Note that the projectors $\Pi_S$ have unit norms and they are orthogonal, namely $\|\Pi_{S_1}\Pi_{S_2}\|=\delta_{S_1,S_2}$. Since each $H_{\text{targ},i}$ is a tensor product of Pauli operators and we assume that the set of $H_{\text{targ},i}$ terms pairwise commute, we have 
\begin{equation}\label{eq:Tr_sum}
\|T_r\|_2=\sum_S\|O_{S,r}\|_2=\sum_S\|O_{S,r}\|_\infty. 
\end{equation}
Because the operator $S$ flips all of the qubits in a certain subset $F\subseteq\{1,2,\cdots,m\}$ of ancilla registers, the operator $O_{S,r}$ is proportional to $\sum_{i\in F}H_\text{targ,$i$}$ (see Equation \ref{eq:htarg} for the definition of $H_{\text{targ},i}$). Therefore the problem of evaluating $\|T_r\|_2$ becomes the problem of finding the coefficients for all of the $O_{S,r}$ operators. These coefficients are precisely given by Algorithm \ref{alg:perturbbound}. We show that in fact the quantities $\gamma_r$ computed at step \ref{step:alpha_r} and $\gamma_{i,r}$ computed at step \ref{step:gamma} satisfy
\begin{equation}\label{eq:gamma_proof}
\begin{array}{ccl}
\gamma_r & = & \|\langle 0^{km}|T_r|0^{km}\rangle\|_\infty \\[0.02in]
\gamma_{i,r} & = & \displaystyle \sum_{\substack{\text{$S$ acting} \\ \text{ on $i$ regs}}}\|\langle 0^{km}|T_rS|0^{km}\rangle\|_\infty.
\end{array}
\end{equation}
Here in the expression for $\gamma_{i,r}$ the summation is over all $S$ operators of the form in Equation \ref{eq:STS_phi} that acts non-trivially on $i$ ancilla registers. The collection of all states of the form $S|0^{km}\rangle$ acting non-trivially on $i$ registers is then the set of all states $|\phi\rangle$, $\phi\in\{0,1\}^{km}$ such that its reduced configuration is $\tilde{\bf c}=(\underbrace{k,\cdots,k}_\text{$i$ times},0,\cdots,0)$. 

We show Equation \ref{eq:gamma_proof} by taking advantage of the hierarchical structure of equivalence classes in the sets $\mathcal{W}_r$, $\mathcal{W}_r^{\bf c}$, and $\widetilde{W}_r^{\bf c}$ (Figure \ref{fig:W_reduce}). 
Observing Equation \ref{eq:Tr_walk} and Equation \ref{eq:t0r_bound} we see that $t_\phi(\phi_0,\phi_1,\cdots,\phi_r)$ is in fact the norm of the $|\phi_0\rangle\langle\phi_r|$ block of $T_r$. For a specific $|\phi_r\rangle=S|0^{km}\rangle$ for some $S$, we could evaluate $\|O_{S,r}\|_2$ by summing over the weight $t_\phi$ of all sequences in $\mathcal{W}_r(0^{km},\phi_r)$. We partition the set $\mathcal{W}_r(0^{km},\phi_r)$ into equivalence classes as discussed before and each equivalence class is associated with an element in $\mathcal{W}_r^{\bf c}({\bf c}_0,{\bf c}_r)$ with ${\bf c}_0$, ${\bf c}_r$ being the configurations of $|0^{mk}\rangle$ and $S|0^{mk}\rangle$ respectively. We could evaluate the weight $t_{\bf c}$ of a configuration sequence $({\bf c}_0,{\bf c}_1,\cdots,{\bf c}_r)\in\mathcal{W}_r^{\bf c}$ as the sum of the weights of all the elements in the equivalent class in $\mathcal{W}_r$ associated with it. Specifically, starting with any sequence $(\phi_0,\cdots,\phi_r)$ in the equivalence class, we could calculate the weight $t_{\bf c}$ by summing over all the local permutations $\zeta$ that produces the elements in the class:
\begin{equation}\label{eq:w_configuration}
t_{\bf c}({\bf c}_0,{\bf c}_1,\cdots,{\bf c}_r)=\sum_{\zeta}t_\phi(\zeta(\phi_0),\zeta(\phi_1),\cdots,\zeta(\phi_r)).
\end{equation}
At each step of a sequence in $\mathcal{W}_r$ going from $\phi_{i-1}$ to $\phi_{i}$, suppose the number of $|1\rangle$ ancilla qubits in the register $p_i$ changes from $x_i$ to $y_i$. To evaluate Equation \ref{eq:w_configuration}, we need to sum over all possible ways in which a $k$-qubit state with Hamming weight $x_i$ can make a transition to a state with Hamming weight $y_i$ through the action of the perturbation $V$. Because local permutations do not change the weight of the sequence $(\phi_0,\cdots,\phi_r)$, the weight $t_{\bf c}({\bf c}_0,\cdots,{\bf c}_r)$ differs from $t_\phi(\phi_0,\cdots,\phi_r)$ by a multiplicative factor. With the specific construction of $V$ in Equation \ref{eq:Vi}, such multiplicative factor can be calculated using $M_{ij}$ in Equation \ref{eq:M}. Let $\Omega_i=M_{x_iy_i}$, which is the number of possible ways for an ancilla register to go from a state with $x_i$ qubits in $|1\rangle$ to one with $y_i$ qubits in $|1\rangle$. Then we have
\begin{equation}
t_{\bf c}({\bf c}_0,{\bf c}_1,\cdots,{\bf c}_r)=\left(\prod_{i=1}^r\Omega_i\right)t_\phi(\phi_0,\phi_1,\cdots,\phi_r)
\end{equation}
with $(\phi_0,\phi_1,\cdots,\phi_r)$ being any sequence with the configuration of $\phi_i$ being ${\bf c}_i$. Similar to Equation \ref{eq:w_configuration}, we could also evaluate the weight of a sequence of reduced configuration by summing over all permutations of $m$ registers:
\begin{equation}\label{eq:w_red_configuration}
\begin{array}{ccl}
\tilde{t}_{\bf c}(\tilde{\bf c}_0,\tilde{\bf c}_1,\cdots,\tilde{\bf c}_r) & = & \displaystyle \sum_{\pi:[m]\mapsto[m]}t_{\bf c}(\pi({\bf c}_0),\pi({\bf c}_1),\cdots,\pi({\bf c}_r)) \\
& = & \displaystyle \prod_{i=1}^{r-1}\frac{1}{|z-E(\tilde{\bf c}_i)|}\prod_{i=1}^r\Omega_i\underbrace{\sum_{\pi:[m]\mapsto[m]}\lambda_{\pi(p_1)}\lambda_{\pi(p_2)}\cdots\lambda_{\pi(p_r)}}_{(*)}.
\end{array}
\end{equation}
Here the term $(*)$ is essentially a monomial symmstric polynomial over the variables $\boldsymbol\lambda=(\lambda_1,\lambda_2,\cdots,\lambda_m)$. We can rewrite it as $m_{{\bf b}_r}(\boldsymbol\lambda)$ with ${\bf b}_r$ being the partition of the symmetric polynomial that keeps track of ``how many registers have been acted on by how many times". For example, if there are three registers that are acted on (i.e.\ have one or more bits flipped in them) once and one register acted on twice, in which case the order of perturbation theory is $r=3\times 1+1\times 2=5$, then ${\bf b}_r=(2,1,1,1)$. In general one could compute ${\bf b}_r$ for a reduced configuration sequence in $\widetilde{\mathcal{W}}_r^{\bf c}$. 

From the arguments so far, it should be clear that Algorithm \ref{alg:walkbound} computes $\tilde{t}_{\bf c}$ correctly. Because for any $S$ of the form in Equation \ref{eq:S_il}, $S|0^{km}\rangle$ is the only state with its configuration and reduced configuration, summing the weights of all sequences $(0^{km},\phi_1,\cdots,\phi_r)$ with $|\phi_r\rangle=S|0^{km}\rangle$ in $\mathcal{W}_r$ is equivalent to summing the weights of all sequences of reduced configurations with $\tilde{\bf c}=(0,0,\cdots,0)$ and $\tilde{\bf c}_r=({k,\cdots,k},0,\cdots,0)$ with the number of elements equal to $k$ being the number of registers that $S$ acts on. Therefore Equation \ref{eq:gamma_proof} holds and the main statement of the theorem is proven. \hfill{$\square$}
%
%

\section{Potential improvement using Schrieffer-Wolff transformation}\label{sec:sw}

Theorem \ref{th:sharpness} presented in Section \ref{subsec:sharp} shows the sharpness of the bounds provided by our algorithm for difference in norm $\|\Sigma_-(z)-H_\text{eff}\|_2$ between the self-energy $\Sigma_-(z)$ and the effective Hamiltonian $H_\text{eff}$ in Equation \ref{eq:sigmaz_heff}. However, the quantity $\|\Sigma_-(z)-H_\text{eff}\|_2$ is itself an upper bound to the actual spectral difference between the gadget and target Hamiltonian (see Theorem \ref{thm:KKR06} as well as numerics in Section \ref{sec:num_ex}) because $\|\Sigma_-(z)-H_\text{eff}\|_2\le\epsilon$ is only a \emph{sufficient} condition that guatantees that the spectral difference is small, namely $\max_j|\lambda_j(\tilde{H})-\lambda_j(H_\text{targ})|\le\epsilon$ with $\lambda_j(H)$ being the $j$-th lowest eigenvalue of $H$. In practice one typically is more concerned about the spectral difference between the target and gadget Hamiltonian than $\|\Sigma_-(z)-H_\text{eff}\|_2$. Therefore to obtain a tighter upper bound to the actual spectral error than the norm difference $\|\Sigma_-(z)-H_\text{eff}\|_2$ based on Feynman-Dyson (FD) series, we need to adopt a different formalism of perturbation theory. Here we consider using the Schrieffer-Wolff (SW) transformation. As we will prove in this section, the SW series generates all the terms in the FD series but also includes more terms that are beyond FD. In Section \ref{sec:num_ex} we numerically show an improved bound for the spectral error based on the SW transformation over $\|\Sigma_-(z)-H_\text{eff}\|_2$.

The Schrieffer-Wolff transformation is a formalism of degenerate perturbation theory where the low energy effective Hamiltonian $H_\text{eff}$ is obtained from the perturbed Hamiltonian by a unitary transformation that makes the perturbed Hamiltonian block diagonal with respect to low and high energy subspaces \cite{BDL11}. Using the same setting as Section \ref{sec:pt}, we consider a perturbed Hamiltonian $\tilde{H}=H+V$ which is a sum of some unperturbed Hamiltonian $H$ that is diagonal in the basis we are assuming and a perturbation $V=V_d+V_{od}$ that contains both diagonal $V_d$ and off-diagonal $V_{od}$ components. The basic idea of SW transformation is to find an anti-Hermitian operator $R$ such that $e^R(H+V)e^{-R}$ is block diagonal with respect to the high energy subspace $\mathcal{L}_+$ and the low energy subspace $\mathcal{L}_-$. The effective low energy Hamiltonian is then $H_\text{eff}=\Pi_-e^R(H+V)e^{-R}\Pi_-$ with $\Pi_-$ being the projector to $\mathcal{L}_-$. 

We denote the adjoint operation of an operator $Y$ on $X$ as $\hat Y(X)=[Y,X]$. Let $\mathcal{O}$ be a super operator that extracts the off-diagonal component of an operator. For example $V_{od}=\mathcal{O}(V)$. Let $\mathcal{K}$ be a super operator such that
\begin{equation}\label{eq:superop_K}
\mathcal{K}(X)=\sum_{i,j}\frac{\langle i|\mathcal{O}(X)|j\rangle}{E_i-E_j}|i\rangle\langle j|
\end{equation}
where the $|i\rangle$, $|j\rangle$ states are the eigenstates of $H$ and the summation is over any $i$, $j$ such that either $|i\rangle\in\mathcal{L}_-$, $|j\rangle\in\mathcal{L}_+$, or $|i\rangle\in\mathcal{L}_+$, $|j\rangle\in\mathcal{L}_-$. 

The anti-Hermitian operator $R$ admits an expansion $R=\sum_{n=1}^\infty R_n$. To ensure that the transformed Hamiltonian $e^{\hat R}(H+V)=e^R(H+V)e^{-R}$ is block diagonal, the $R_n$ terms are given by\footnote{In \cite{BDL11} the authors consider a setting where $\tilde{H}=H+\varepsilon V$ and $R$ is a Taylor expansion $R=\sum_{n=1}^\infty R_n\varepsilon^n$. Here we absorb the $\varepsilon$ parameter into $V$ and replace $\varepsilon$ with 1.}
\begin{equation}\label{eq:Rn}
\begin{array}{ccl}
R_1 & = & \mathcal{K}(V_{od}) \\
R_2 & = & \mathcal{K}({\hat R}_1V_d) \\
R_n & = & -\mathcal{K}{\hat V}_d(R_{n-1})+\sum_{j\ge 1}a_{2j}\mathcal{K}{\hat R}^{2j}(V_{od})_{n-1}
\end{array}
\end{equation}
where $a_m=\frac{2^m}{m!}B_m$, $B_m$ being the $m$-th Bernoulli number and
\begin{equation}\label{eq:Rkm}
{\hat R}^k(V_{od})_m=\sum_{\substack{n_1+\cdots+n_k=m \\ n_1,\cdots,n_k\ge 1}}{\hat R}_{n_1}{\hat R}_{n_2}\cdots{\hat R}_{n_k}(V_{od}).
\end{equation}
The effective Hamiltonian $H_{\text{eff,SW}}$ is then given by
\begin{equation}
H_\text{eff,SW}=H_-+V_-+\sum_{r=2}^\infty\underbrace{\sum_{j\ge 1}b_{2j-1}\Pi_-{\hat R}^{2j-1}(V_{od})_{r-1}\Pi_-}_{H_{\text{eff},r}}
\end{equation}
where the coefficients $b_n=\frac{2(2^{2n}-1)}{(2n)!}B_{2n}$. Note that the summation in $H_{\text{eff},r}$ over $j\ge 1$ is not infinite, because from Equation \ref{eq:Rkm} we see that ${\hat R}^k(V_{od})_m=0$ if $k>m$. From the definition of $H_{\text{eff},r}$ it is clear that for any $r$, the $r$-th order effective Hamiltonian $H_{\text{eff},r}$ must contain a term of the form
\begin{equation}\label{eq:specialterm}
-b_1\Pi_-{\hat V}_{od}(-\mathcal{K}{\hat V}_d)^{r-2}R_1\Pi_-.
\end{equation}
There are of course other terms appearing at any order $r$ and in \cite{BDL11} the authors have created an elegant diagrammatic technique for enumerating the terms. However, here we focus on the terms of the form in \eqref{eq:specialterm} (which correspond to tree diagrams that are simply a linear chain of nodes) and show that at any order $n$, in some sense \eqref{eq:specialterm} is equivalent to the entire $r$-th order term $T_r$ in the FD series (Equations \ref{eq:sigmaz} and \ref{eq:T_r}). We state it precisely in the following theorem.
\begin{theorem}\label{thm:sw_fd}
If $H$ has a unique eigenvalue $E_0$ that is below the cutoff between low and high energy subspaces $\mathcal{L}_-$ and $\mathcal{L}_+$, then for any $r\ge 2$
\begin{equation}\label{eq:TrE0}
T_r(E_0)=-b_1\Pi_-{\hat V}_{od}(-\mathcal{K}{\hat V}_d)^{r-2}R_1\Pi_-.
\end{equation}
Here we write $T_r$ explicitly as a function of $z$, namely $T_r(z)=V_{-+}(G_+(z)V_+)^{r-2}G_+V_{+-}$.
\end{theorem}
$\quad$\\
For simplicity, we denote the eigenstates of $H$ as $|i\rangle$ with $H|i\rangle=E_i|i\rangle$. To prove the statement we first show that for any $r\ge 1$,
\begin{equation}\label{eq:KVdr}
\begin{array}{l}
\displaystyle(\mathcal{K}{\hat V}_d)^rR_1=\sum_{i\in\mathcal{L}_-}\sum_{j_1\in\mathcal{L}_+}\cdots\sum_{j_{r+1}\in\mathcal{L}_+}\langle i|V|j_1\rangle\frac{1}{E_i-E_{j_1}}\langle j_1|V|j_2\rangle\cdots \\
\displaystyle\makebox[1in]{}\cdots\langle j_r|V|j_{r+1}\rangle\left(\frac{1}{E_{j_{r+1}}-E_i}|j_{r+1}\rangle\langle i|+\frac{1}{E_i-E_{j_{r+1}}}|i\rangle\langle j_{r+1}|\right).
\end{array}
\end{equation}
We prove \eqref{eq:KVdr} inductively on $r$. The base case is $r=1$, By straightforward calculation
\begin{equation}
\mathcal{K}{\hat V}_dR_1=\sum_{i\in\mathcal{L}_-}\sum_{j\in\mathcal{L}_+}\sum_{k\in\mathcal{L}_+}\langle i|V|j\rangle\frac{1}{E_i-E_j}\langle j|V|k\rangle\left(\frac{1}{E_k-E_i}|k\rangle\langle i|+\frac{1}{E_i-E_k}|i\rangle\langle k|\right).
\end{equation}
The case for general $k$ can be proved by similar calculations using the definitions of $\mathcal{K}$ and $V_d$. 

$\quad$\\
\noindent{\bf Proof of Theorem \ref{thm:sw_fd}.} Rewriting the projected operators $V_{-+}$, $G_+$ etc into a summation over $|i\rangle\langle j|$ blocks, we have for example
\begin{equation}
V_{-+}=\sum_{i\in\mathcal{L}_-}\sum_{j\in\mathcal{L}_+}\langle i|V|j\rangle|i\rangle\langle j|,\qquad
G_+(z)=\sum_{i\in\mathcal{L}_+}\frac{1}{z-E_i}|i\rangle\langle i|
\end{equation}
and similar for $V_+$ and $V_{+-}$. Hence we could rewrite $T_r(z)$ as
\begin{equation}\label{eq:VGVGV}
\begin{array}{l}
\displaystyle V_{-+}(G_+(z)V_+)^{r-2}G_+(z)V_{+-}=\sum_{i\in\mathcal{L}_-}\sum_{j_1\in\mathcal{L}_+}\cdots\sum_{j_{r-1}\in\mathcal{L}_+}\sum_{\ell\in\mathcal{L}_-}\langle i|V|j_1\rangle\frac{1}{z-E_{j_1}}\langle j_1|V|j_2\rangle\cdots \\
\displaystyle\makebox[2in]{}\cdots\frac{1}{z-E_{j_{r-1}}}\langle j_{r-1}|V|\ell\rangle|i\rangle\langle \ell|.
\end{array}
\end{equation}
Using Equation \ref{eq:KVdr} and $\Pi_-=\sum_{i\in\mathcal{L}_-}|i\rangle\langle i|$, we have
\begin{equation}\label{eq:b1Pm}
\begin{array}{l}
\displaystyle -b_1\Pi_-{\hat V}_{od}(-\mathcal{K}{\hat V}_d)^{r-2}R_1\Pi_-=\frac{1}{2}\sum_{i\in\mathcal{L}_-}\sum_{j_1\in\mathcal{L}_+}\cdots\sum_{j_{r-1}\in\mathcal{L}_+}\sum_{\ell\in\mathcal{L}_-}\langle i|V|j_1\rangle\frac{1}{E_i-E_{j_1}}\langle j_1|V|j_2\rangle\cdots \\
\displaystyle \makebox[2in]{}\cdots\frac{1}{E_i-E_{j_{r-1}}}(|\ell\rangle\langle i|+|i\rangle\langle\ell|).
\end{array}
\end{equation}
Comparing Equations \ref{eq:b1Pm} with \ref{eq:VGVGV} and the main equation \eqref{eq:TrE0} follows. \hfill{$\square$}

\section{Numerical example}\label{sec:num_ex}

\subsection{\textsc{PerturbBound} vs.\ Simple upper bound}\label{sec:num_pt}

\begin{figure}
\begin{center}
\includegraphics[scale=0.8]{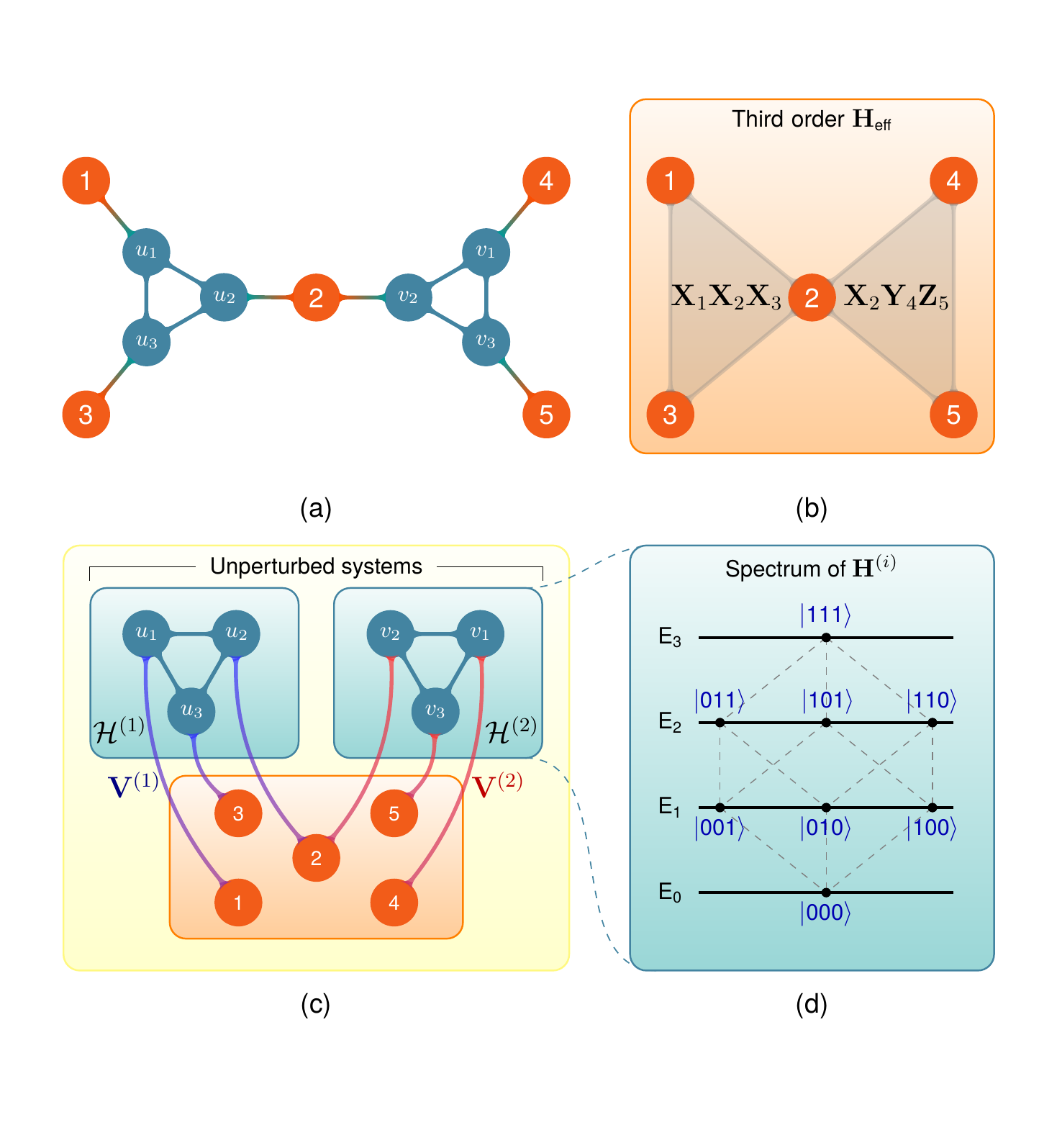}
\fcaption{A numerical example for demonstrating our algorithm estimating the perturbative error. (a) The 11-spin gadget Hamiltonian. Each node corresponds to a spin-1/2 particle and each edge represents an interaction term in the Hamiltonian between two spins. (b) The target 3-body Hamiltonian $H_\text{eff}=\alpha_1X_1X_2X_3+\alpha_2X_2Y_4Z_5$. (c) Rearranging and partitioning the system in (a) according to the setting of perturbation theory used. Here each unperturbed system $H^{(i)}$ consists of three ferromagnetically interacting spins. (d) Spectrum of each subsystem $H^{(i)}$ in (a), $i\in\{1,2\}$. Here each node represents an eigenstate of $H^{(i)}$. Nodes on a same horizontal dashed line belong to the same energy subspace $\mathcal{P}_j$. There is an edge $(\phi_1,\phi_2)$ iff $\|\langle \phi_1|V|\phi_2\rangle\|\neq 0$. For example, if we consider this diagram as representing $H^{(1)}$, since $V^{(1)}|001\rangle_{u_1u_2u_3}\propto(|101\rangle+|011\rangle+|000\rangle)_{u_1u_2u_3}$ we connect the $|001\rangle$ with the nodes representing $|101\rangle$, $|011\rangle$ and $|000\rangle$.}
\label{fig:num_ex_big}
\end{center}
\end{figure}

Here we compare the tightness of bounds obtained by \textsc{PerturbBound} and simple upper bounds (from the right hand side of Equation \ref{eq:Tr}). Consider applying the gadget construction in Section \ref{sec:pt} on the 3-body target Hamiltonian is $H_\text{eff}=\alpha_1X_1X_2X_3+\alpha_2X_2Y_4Z_5$ where $\alpha_1$ and $\alpha_2$ are real coefficients (Figure \ref{fig:num_ex_big}b). The resulting gadget Hamiltonian is described in Figure \ref{fig:num_ex_big}a, which can be expressed in form of the general setting $\tilde{H}=H+V$. Here the unperturbed Hamiltonian $H$ and perturbation $V$ are defined as

\begin{equation}\label{eq:bitflip}
\begin{array}{ll}
H = H^{(1)} + H^{(2)}, & \displaystyle\qquad\qquad H^{(1)}=\frac{\Delta}{4}(3{\bf I}-Z_{u_1}Z_{u_2}+Z_{u_2}Z_{u_3}+Z_{u_1}Z_{u_3}) \\[0.1in]
& \displaystyle\qquad\qquad H^{(2)}=\frac{\Delta}{4}(3{\bf I}-Z_{v_1}Z_{v_2}+Z_{v_2}Z_{v_3}+Z_{v_1}Z_{v_3}) \\[0.1in]
V = V^{(1)} + V^{(2)}, & \qquad\qquad V^{(1)}=\mu_1(X_1X_{u_1}+X_2X_{u_2}+X_3X_{u_3}) \\[0.05in]
& \qquad\qquad V^{(2)}=\mu_2(Y_4X_{v_1}+X_2X_{v_2}+Z_5X_{v_3})
\end{array}
\end{equation}
where spins with $u_i$ and $v_i$ labels belong to the two unperturbed subsystems. Here we let $\Delta$ be orders of magnitude larger than $\mu_1$ and $\mu_2$ and keep the coefficients $\mu_1$ and $\mu_2$ as
\begin{equation}
\mu_1 = \left(\frac{\alpha_1\Delta^2}{6}\right)^{1/3},\qquad
\mu_2 = \left(\frac{\alpha_2\Delta^2}{6}\right)^{1/3}
\end{equation}
where $\alpha_1$ and $\alpha_2$ are parameters related to the low energy effective Hamiltonian (see Equation \ref{eq:heff_ex}).
In Figure \ref{fig:num_ex_big}c we explicitly partition the Hamiltonian in terms of $H$ and $V$.

The low-energy subspace of the total Hamiltonian $\tilde{H}$ is then $\mathcal{L}_-=\mathcal{L}_-^{(1)}\otimes\mathcal{L}_-^{(2)}$. Inspecting the expressions $H^{(1)}$ and $H^{(2)}$ gives the low energy subspaces for each subsystem: $\mathcal{L}_-^{(1)}=\text{span}\{|000\rangle_{u_1u_2u_3},|111\rangle_{u_1u_2u_3}\}$ and $\mathcal{L}_-^{(2)}=\text{span}\{|000\rangle_{v_1v_2v_3},|111\rangle_{v_1v_2v_3}\}$. For each subsystem $i\in\{1,2\}$, the subspaces of $H^{(i)}$ and their corresponding energies are

\begin{equation}
\begin{array}{ll}
\mathcal{P}_0 = \text{span}\{|000\rangle\}, & E_0 = 0 \\
\mathcal{P}_1 = \text{span}\{|001\rangle,|010\rangle,|100\rangle\}, & E_1 = \Delta \\
\mathcal{P}_2 = \text{span}\{|011\rangle,|101\rangle,|110\rangle\}, & E_2 = \Delta \\
\mathcal{P}_3 = \text{span}\{|111\rangle\}, & E_3 = 0.
\end{array}
\end{equation}

In Figure \ref{fig:num_ex_big}d we show the spectrum of each subsystem. The matrix $M$ defined in Equation \ref{eq:M} is also involved in the computation of the upper bound to $\|T_r\|$. We could interpret $M$ from Figure \ref{fig:num_ex_big}d. One could regard $M_{ij}$ as the maximum, over all eigenstates of $H$ in $\mathcal{P}_i$, number of possible transitions from a particular $|u\rangle\in\mathcal{P}_i$ to an eigenstate in $\mathcal{P}_j$. Precisely,
\begin{equation}
M_{ij}=\max_{|u\rangle\in\mathcal{P}_i}Card\{|v\rangle\in\mathcal{P}_j|\|\langle v|V|u\rangle\|\neq 0\}
\end{equation}
where $Card\{\cdot\}$ stands for cardinality (number of distinct elements) of a set.
We could then determine that
\begin{equation}\label{eq:M_ex}
{M}=
\kbordermatrix{
  & \mathcal{P}_0 & \mathcal{P}_1 & \mathcal{P}_2 & \mathcal{P}_3 \\
\mathcal{P}_0 &  & 3 &  &  \\
\mathcal{P}_1 & 1 &  & 2 &  \\
\mathcal{P}_2 &  & 2 &  & 1 \\
\mathcal{P}_3 &  &  & 3 & 
}
\end{equation}
where the row and column indices start from 0 because the subspaces $\mathcal{P}_0$, $\mathcal{P}_1$, $\cdots$, have indices that start from 0. 

From Figure \ref{fig:num_ex_big}a and \ref{fig:num_ex_big}c we can see that the unperturbed system $H$ essentially consists of two identical 4-level systems with energy levels $E_0$, $E_1$, $E_2$ and $E_3$. This gives rise to in total 9 possible energy combinations.

With the matrix ${M}$ worked out as in Equation \ref{eq:M_ex}, we could use the algorithm \textsc{WalkBound} in Section \ref{subsec:algo} to find a tight upper bound for $\|T_r\|_\infty$ at any order $r$. After a certain order $p$, when the upper bound becomes less than the tolerance $10^{-8}$, we use Equation \ref{eq:Tr} to bound the terms from $p+1$ to infinity.

\begin{figure}
\begin{center}
\includegraphics[scale=0.8]{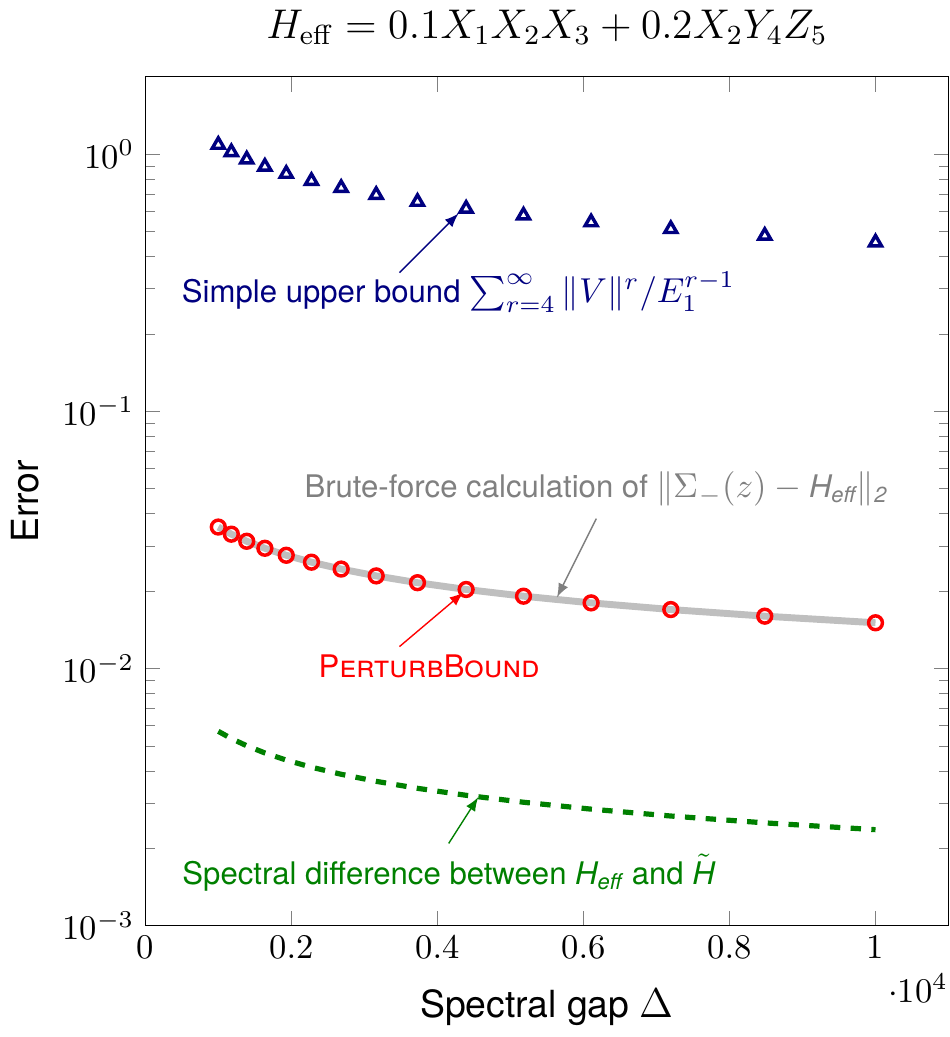}
\fcaption{Comparison between the upper bounds computed using the \textsc{PerturbBound} and the norm computed using (inefficient) explicit matrix-matrix multiplication. The ``actual spectral error'' in this plot shows the maximum difference between the eigenvalues of $H_\text{eff}$ and their counterparts in $\tilde{H}$, which are the energies of its $2^N$ lowest eigenstates with $N=5$ being the number of particles that $H_\text{eff}$ acts on (Figure \ref{fig:num_ex_big}b). The actual spectral error is always lower than the error computed based on $\|\Sigma_-(z)-H_\text{eff}\|_2$ because $\|\Sigma_-(z)-H_\text{eff}\|_2\le\epsilon$ is only a \emph{sufficient} condition that guarantees the spectral difference between $\tilde{H}$ and $H_\text{eff}$ being within $\epsilon$ (see \cite[Theorem 1]{CK16}).}
\label{fig:num_plot}
\end{center}
\end{figure}

Using the perturbation series in Equation \eqref{eq:sigmaz} we could show that if we truncate the series at the 3rd order, namely $\Sigma_-(z)=H_\text{eff}+T_4+T_5+\cdots$, we have the effective 3-body Hamiltonian 
\begin{equation}\label{eq:heff_ex}
H_\text{eff}=\alpha_1 X_1X_2X_3+\alpha_2 X_2Y_4Z_5+\gamma I
\end{equation}
with $\gamma$ being the magnitude of the spectral shift. Here we let $\alpha_1=0.1$ and $\alpha_2=0.2$. Then the entire Hamiltonian $\tilde{H}=H+V$ in Equation \ref{eq:bitflip} is only dependent on a free parameter $\Delta$. In order to test our algorithm for bounding perturbative terms, we treat terms from 4th order onward as errors in the perturbation series. This amounts to estimating $\|\Sigma_-(z)-H _\text{eff}\|_2$. We could compute this value by explicitly computing $\Sigma_-(z)$ by its definition $z{I}-(\tilde{G}_-(z))^{-1}$ and then evaluating $\|\Sigma_-(z)-H_\text{eff}\|_2$. This method is inefficient since it requires inverting an exponentially large matrix with respect to system size, but yields an accurate estimation for the error $\|\Sigma_-(z)-H_\text{eff}\|_2$. We will use it as a benchmark for comparison with the upper bound computed by the new algorithm developed here. As shown Figure \ref{fig:num_plot}, the upper bounds computed by \textsc{PerturbBound} are tight with respect to the exact calculation. For the purpose of comparison we also compute the error bound due to triangle inequality (see Equation \ref{eq:Tr}). We explicitly computed $\|V\|_2$ (while in practice one may use some upper bound for $\|V\|_2$ which could loosen the bound further but here for comparison we use the exact value) and bounded $\|G_+\|_2$ from above by $1/E_1$. Hence the simple bound of error terms from a certain order to infinity based on Equation \ref{eq:Tr} becomes $\sum_{r=4}^\infty\|V\|_2^r/E_1^{r-1}=\|V\|_2^4/(E_1^2(E_1-\|V\|_2))$. When implementing our algorithm for the numerical example concerned in this section, we compute $\tau_r=\textsc{PerturbBound}(r,\boldsymbol\lambda,{M})$ for $r$ from 4 to a value $p$ such that $\tau_{p}\le 10^{-20}$. Then we resort to Equation \ref{eq:Tr} for computing an upper bound to $\|T_{p+1}+T_{p+2}+\cdots\|_2$.

\begin{figure}
\begin{center}
\includegraphics[scale=0.8]{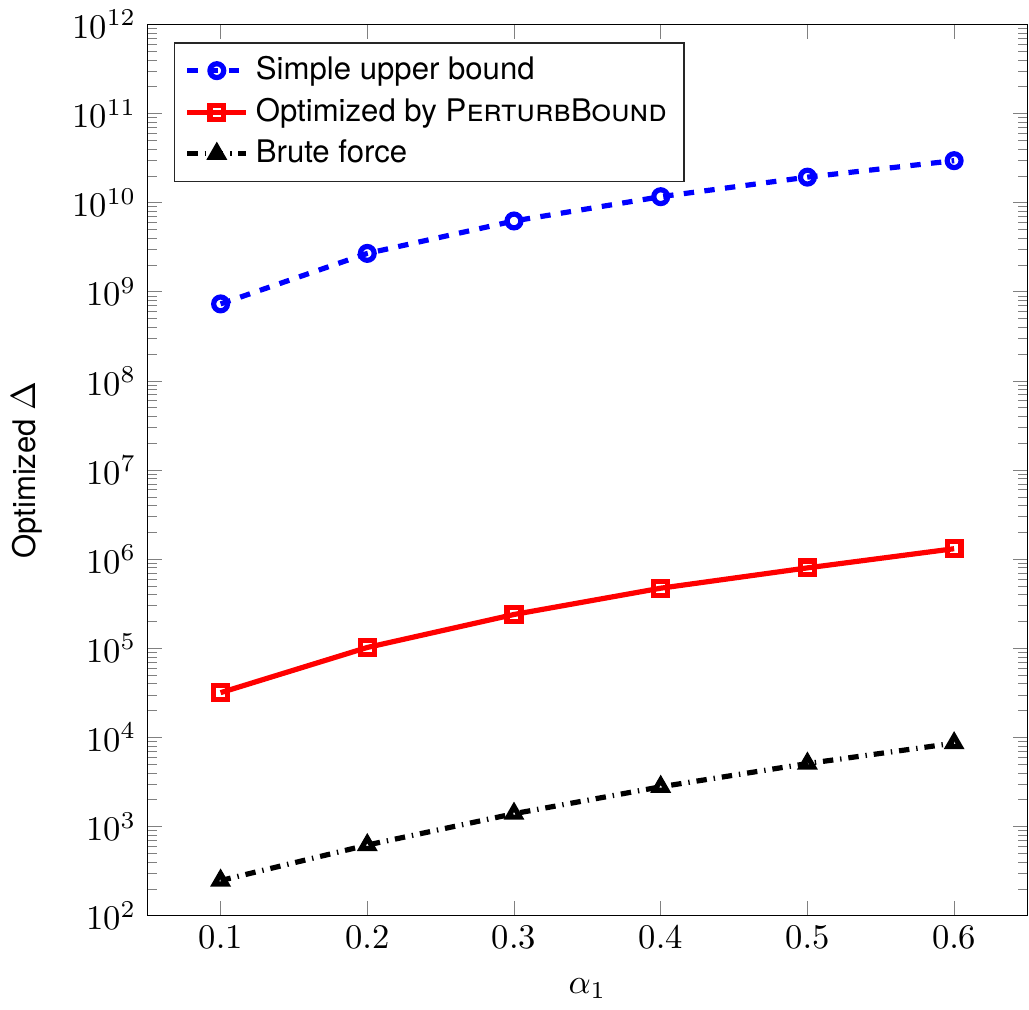}
\fcaption{Comparison between the optimized $\Delta$ based on crude error bounds using geometric series (``simple upper bound" in Figure \ref{fig:num_plot}), the optimized $\Delta$ based on the \textsc{PerturbBound} algorithm presented in Section \ref{subsec:algo} and optimized $\Delta$ based on spectral error between the target and the gadget Hamiltonian computed by brute-force diagonalizing both Hamiltonians. Here we use the target Hamiltonian in Figure \ref{fig:num_ex_big}b with $\alpha_2=0.2$ fixed and $\alpha_1$ varying from 0.1 to 0.6. It can be observed that our algorithm significantly improves the assignments for $\Delta$. The gap between the brute force case and the \textsc{PerturbBound} case is due to the fact that $\|\Sigma_-(z)-H_\text{eff}\|_2\le\epsilon$ is only a \emph{sufficient} condition that guarantees the spectral error to be within $\epsilon$.}
\label{fig:num_ex2}
\end{center}
\end{figure}

The ultimate purpose for finding tight error bound in the perturbation theory is to find lower assignments to $\Delta$ while maintaining the spectral error between the target and the gadget Hamiltonian within $\epsilon$. As mentioned in Section \ref{sec:improve}, with an algorithm for computing an upper bound to the spectral error we could find the optimal $\Delta$ assignment based on this algorithm by using binary search to find a $\Delta$ such that the error bound is $\epsilon$. In Figure \ref{fig:num_ex2} we show the result of implementing such binary search for three means for estimating the spectral error: 1) crude upper bound based on geometric series described in Equation \ref{eq:Tr}; 2) upper bound computed using the algorithms presented in Section \ref{subsec:algo}; 3) brute-force diagonalization of both the target and gadget Hamiltonian to get the exact eigenvalues. The third option is impractical for general quantum systems of many qubits due to the exponential size of the Hilbert space, though it provides the exact spectral error. The first option is computationally trivial but yields extremely large assignments of $\Delta$ (Figure \ref{fig:num_ex2}). Our algorithm strikes a balance between the two cases by avoiding intense computation while generating $\Delta$ assignments that are orders of magnitude more practical than the first alternative. 

\subsection{Error bounds based on Feynman-Dyson (FD) and Schrieffer-Wolff (SW) series}\label{subsec:sw_fd}

In Section \ref{sec:sw} we showed that the Schrieffer-Wolff expansion includes more terms than Feynman-Dyson series. Could one potentially improve estimation on the spectral error by adopting Schrieffer-Wolff instead of Feynman-Dyson formalism? Here we show numerical evidence that one indeed could signaficantly improve the error bound. Consider an example where the target Hamiltonian is $H_\text{targ}=\alpha X_1X_2X_3$ with $\alpha=0.1$ and the gadget Hamiltonian $\tilde{H}=H+V$ is constructed by adding three ancilla qubits $u_1$, $u_2$ and $u_3$ and defining the Hamiltonians as the following: 
\begin{equation}\label{eq:HV_ex2}
\begin{array}{l}
\displaystyle H=\frac{\Delta}{4}(3{\bf I}-Z_{u_1}Z_{u_2}-Z_{u_1}Z_{u_3}-Z_{u_2}Z_{u_3}) \\
\displaystyle V=\mu(X_1X_{u_1}+X_2X_{u_2}+X_3X_{u_3}).
\end{array}
\end{equation}
where $\mu=\left(\frac{\alpha\Delta^2}{6}\right)^{1/3}$.
Because $H_\text{targ}$ is 3-body our effective Hamiltonian is truncated at the third order and the remaining terms in the expansion are considered as error:
\begin{equation}\label{eq:FD_ex2}
\begin{array}{ccl}
\Sigma_-(z) & = & \underbrace{\frac{3\mu^2}{z-\Delta}\Pi_-+\frac{\Delta^2}{(z-\Delta)^2}\alpha X_1X_2X_3\otimes(|000\rangle\langle 111|_{u_1u_2u_3}+|111\rangle\langle 000|_{u_1u_2u_3})}_{H_\text{eff}=T_1(z)+T_2(z)+T_3(z)} \\
& + & T_4(z)+T_5(z)+\cdots.
\end{array}
\end{equation} 
Applying Schrieffer-Wolff transformation to the gadget Hamiltonian yields the low-energy effective Hamiltonian
\begin{equation}\label{eq:SW_ex2}
H_\text{eff,SW} = \underbrace{b_1\Pi_-{\hat R}_1(V_{od})\Pi_-}_{\text{$2^\text{nd}$ order}}
+\underbrace{b_1\Pi_-{\hat R}_2(V_{od})\Pi_-}_{\text{$3^\text{rd}$ order}}
+\underbrace{b_1\Pi_-{\hat R}_3(V_{od})\Pi_-+b_3\Pi_-{\hat R}_1^3(V_{od})\Pi_-}_{\text{$4^\text{th}$ order}}
+\cdots.
\end{equation}
Because the ground state energy of the unperturbed Hamiltonian is 0, the zeroth order term in the expansion \eqref{eq:SW_ex2} vanishes. From Equation \ref{eq:HV_ex2} the projection of $V$ in the low energy subspace $\mathcal{L}_-$ is 0, thus the first order term also vanishes. The second order term could be rearranged as $-b_1\Pi_-{\hat V}_{od}R_1\Pi_-$, which according to Theorem \ref{thm:sw_fd} is equivalent to the second order term in the Feynman-Dyson series in Equation \ref{eq:FD_ex2} for $z\rightarrow 0$. At third order, Schrieffer-Wolff expansion gives $b_1\Pi_-{\hat V}_{od}\mathcal{K}{\hat V}_dR_1\Pi_-$. Applying Theorem \ref{thm:sw_fd} with $r=3$ we see that this is equivalent to the third order term in the Feynman-Dyson series. Hence up to third order, both formalisms of perturbation theory match up. However, at the fourth order, which is the leading term for the error, difference between the two formalisms starts to show. From the recursive relationship for $R_n$ in Equation \ref{eq:Rn} we see that $R_3$ contains a term $-\mathcal{K}{\hat V}_d(R_2)$. So the fourth order term in Equation \ref{eq:SW_ex2} must contain a term
\begin{equation}
-b_1\Pi_-{\hat V}_{od}(-\mathcal{K}{\hat V}_d(R_2))\Pi_-=-b_1\Pi_-{\hat V}_{od}(-\mathcal{K}{\hat V}_d)^2R_1\Pi_-,
\end{equation}
which is equivalent to the entire fourth-order term of the Feynman-Dyson series (Theorem \ref{thm:sw_fd} with $r=4$). The other terms at the fourth order in Equation \ref{eq:SW_ex2} are beyond Feynman-Dyson series. For example the second term at the fourth order $b_3\Pi_-{\hat R}_1^3(V_{od})\Pi_-$ corresponds  to virtual transitions that switches between $\mathcal{L}_-$ and $\mathcal{L}_+$ multiple times. This violates conditions \ref{cond:Lm} and \ref{cond:Lp} in Section \ref{subsec:reduce} for sequences $(\phi_0,\cdots,\phi_r)$ that contribute non-trivially to $T_r$, which results in such terms being excluded from the Feynman-Dyson series.

\begin{figure}
\begin{center}
\includegraphics[scale=0.8]{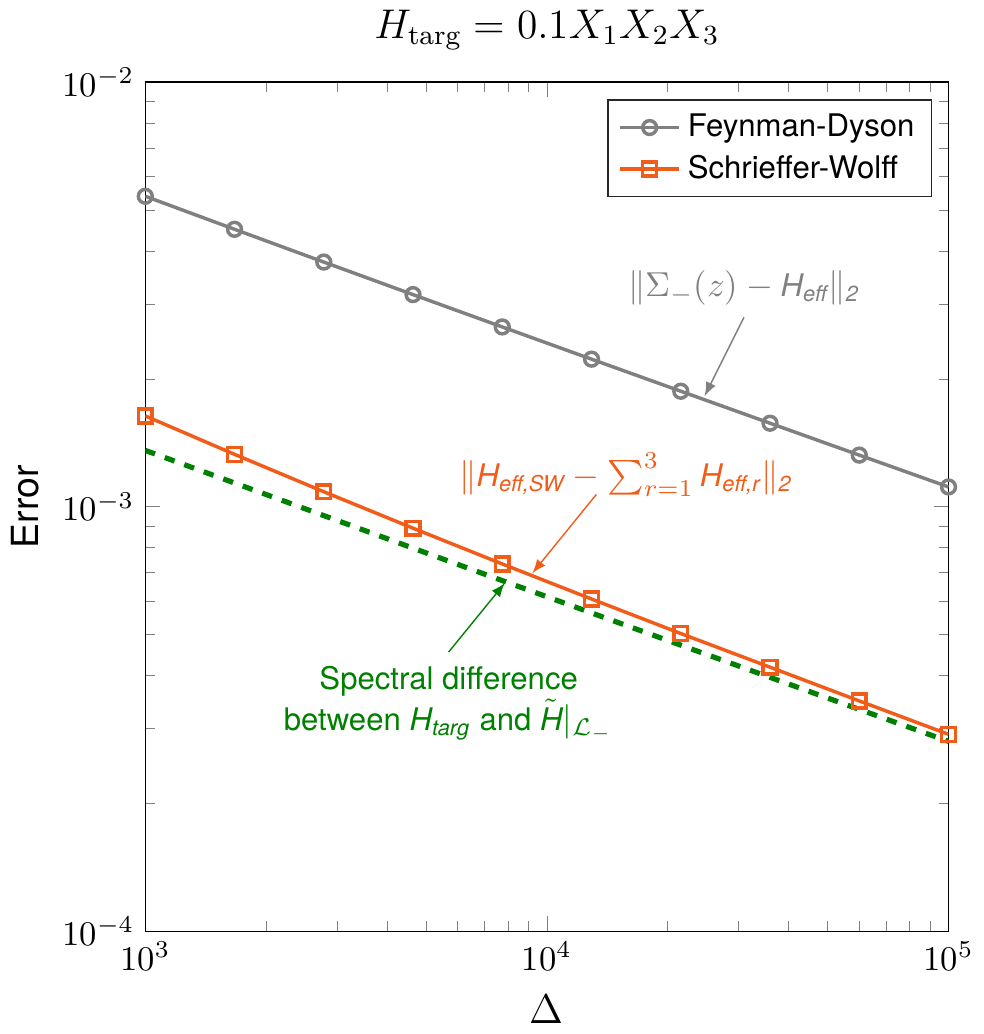}
\fcaption{Comparison between the error bounds computed based on Feynman-Dyson series (Section \ref{sec:errorbound}) and Schrieffer-Wolff transformation (Section \ref{sec:sw}). Here we also show the maximum difference between the lowest $2^3=8$ energy levels of the gadget Hamiltonian and the corresponding level of the target Hamiltonian.}
\label{fig:sw_fd}
\end{center}
\end{figure}

For varying values of $\Delta$, we calculate the error estimates based on both formulations of perturbation theory and compare them in Figure \ref{fig:sw_fd}. We have also explicitly diagonalized the target and gadget Hamiltonian and plotted the difference between the low-lying energy levels. The results in Figure \ref{fig:sw_fd} shows that Schrieffer-Wolff perturbation theory clearly yields tighter error bounds. The error bounds using Schrieffer-Wolff transformation in Figure \ref{fig:sw_fd} are computed by explicit enumeration and evaluation of the terms in the perturbative expansion following Section \ref{sec:sw}, which is clearly not scalable due to the exponential size of the Hilbert space. The algorithms that we have developed in Section \ref{subsec:algo} could efficiently bound only a subset of the terms in the Schrieffer-Wolff series, namely those of the form in Theorem \ref{thm:sw_fd}. Bounding the remaining terms in the Schrieffer-Wolff series with similar effectiveness as our algorithms for Feynman-Dyson series requires additional insight and is beyond the scope of our present study.

\section{Summary and conclusion}

Perturbative gadgets are the only technique available (as of now and as far as the authors are concerned) for reducing arbitrary many-body Hamiltonian to two-body ones. One of the disadvantages of this technique is the large energy gap $\Delta$ needed in the construction of the gadget Hamiltonian, rendering it unnatural in the context of physical systems. Here we address this issue by considering the optimization problem of finding the minimum value of $\Delta$ that yields error no greater than a prescribed threshold $\epsilon$ (Figure \ref{fig:flowchart}). A crucial component of this optimization program is to find tight upper bounds to error terms arising at arbitrary order perturbation theory. In this sense our work is a generalization of \cite{CRBK14} to include the gadget constructions in \cite{KKR06,JF08}. 

The problem of computing the error exactly is hard in general because of the exponential size of the Hilbert space. Alternatively, crude upper bounds are trivially attainable via for instance submultiplicativity of operators ($\|AB\|\le\|A\|\cdot\|B\|$). These bounds are hardly useful for the purpose of \emph{optimizing} the gadget parameters. However, by exploiting the structure of the Hamiltonian we are able to find error bounds that are both orders of magnitude tighter than the crude alternatives (Section \ref{sec:num_pt}) and efficiently computable (Section \ref{subsec:algo}). Each term in the perturbative expansion at a given order is a summation of exponentially many terms. We start from reducing the size of the set of summation from exponential to polynomial in the number of ancilla registers by taking advantage of the structure in the perturbation. We show that there is a hierarchy of equivalence classes (Section \ref{subsec:reduce}) that allows us to accomplish the reduction. The algorithms for computing the error bounds presented in Section \ref{subsec:algo} take advantage of such hierarchical structure. In the special case where the target terms $H_{\text{targ},i}$ pairwise commute, we show that our error bounds are sharp (Section \ref{subsec:sharp}). 

However, a gap still exists (Figure \ref{fig:num_ex2}) between the output of our algorithm and the result of brute-force optimization. This gap is due to the machinery of perturbation theory that we use (illustrated in Figure \ref{fig:num_plot}), which is based on the Feynman-Dyson series. In Section \ref{subsec:sw_fd} we observe numerically that using the Schrieffer-Wolff transformation \cite{PhysRev.149.491,BDL11} instead may enable one to get closer to the brute-force results (Figure \ref{fig:sw_fd}). This improvement may be explained by Theorem \ref{thm:sw_fd} which says that a specific class of terms in the Schrieffer-Wolff series already  captures all of the terms in the Feynman-Dyson series (Section \ref{sec:sw}). It is tempting to consider whether our technique can be applied to obtain efficient error bounds for Schrieffer-Wolff series. One challenge in this regard is that our efficient algorithm is built on the observation that the terms at each order is essentially a summation of walks in the eigenspace of the unperturbed Hamiltonian, per Equation \ref{eq:Tr_walk}. This combinatorial picture of summing over walks comes from the matrix product structure of the self-energy expansion (Equation \ref{eq:sigmaz}). Whether this same structure exists in Schrieffer-Wolff transformation (and other formalisms of perturbation theory) remains to be assessed.  

\section{Acknowledgments}

Y.\ C.\ would like to acknowledge financial support from Dimitris N.\ Chorafas Foundation, as well as Qatar Energy and Environmental Research Institute (QEERI) for accomodating a visit, during which this work was partially completed. The authors would like to thank the anonymous reviewers for their helpful comments.


\bibliographystyle{unsrt}
\bibliography{gadgetReferences}

%
\appendix{: An example for illustrating notions introduced in Section \ref{subsec:reduce}}\label{subsec:example}

This example is essentially the one considered in Section \ref{sec:num_ex} but here we abstract out only the revelant aspects of the example without going into full detail. Suppose our target Hamiltonian $H_\text{targ}$ is a sum of two 3-local terms that need to be reduced to 2-local using the gadget construction (Section \ref{sec:pt}). Our gadget Hamiltonian $\tilde{H}=H+V$ has the unperturbed part $H=H^{(1)}+H^{(2)}$ acting on two registers of three ancilla qubits (because the target terms are 3-local). The perturbation $V$ couples to each register of ancillas with interaction strengths $\lambda_1$ and $\lambda_2$ (as a reminder, see Equation \ref{eq:Vi} and the restriction that $\lambda_{i,j}=\lambda_i$ introduced at the beginning of Section \ref{subsec:reduce}). Hence $m=2$ and $k=3$ in this example and the low energy level of $H$ satisfies $E(\phi)=E^{(1)}(j_1)+E^{(2)}(j_2)=0$ with $j_1$, $j_2$ being either 0 or 3. At second order, from previous discussion we see that the sequences of reduced configurations that contribute non-trivially to $T_2$ are
\begin{equation}\label{eq:red_c_seq}
\begin{array}{c}
\tilde{\bf c}_0=
\begin{pmatrix}0 \\ 0\end{pmatrix}\rightarrow
\tilde{\bf c}_1=
\begin{pmatrix}0 \\ 1\end{pmatrix}\rightarrow
\tilde{\bf c}_2=
\begin{pmatrix}0 \\ 0\end{pmatrix} \\[0.15in]
\tilde{\bf c}_0=
\begin{pmatrix}0 \\ 3\end{pmatrix}\rightarrow
\tilde{\bf c}_1=
\begin{pmatrix}1 \\ 3\end{pmatrix}\rightarrow
\tilde{\bf c}_2=
\begin{pmatrix}0 \\ 3\end{pmatrix} \\[0.15in]
\tilde{\bf c}_0=
\begin{pmatrix}0 \\ 3\end{pmatrix}\rightarrow
\tilde{\bf c}_1=
\begin{pmatrix}0 \\ 2\end{pmatrix}\rightarrow
\tilde{\bf c}_2=
\begin{pmatrix}0 \\ 3\end{pmatrix} \\[0.15in]
\makebox[0.07in]{}
\tilde{\bf c}_0=
\begin{pmatrix}3 \\ 3\end{pmatrix}\rightarrow
\tilde{\bf c}_1=
\begin{pmatrix}2 \\ 3\end{pmatrix}\rightarrow
\tilde{\bf c}_2=
\begin{pmatrix}3 \\ 3\end{pmatrix}.
\end{array}
\end{equation}
Accordingly, the set $\mathcal{W}_2^{\bf c}$ consists of the following sequences of configurations
\begin{equation}\label{eq:c_seq}
\begin{array}{cc}
{\bf c}_0=
\begin{pmatrix}0 \\ 0\end{pmatrix}\rightarrow
{\bf c}_1=
\begin{pmatrix}1 \\ 0\end{pmatrix}\rightarrow
{\bf c}_2=
\begin{pmatrix}0 \\ 0\end{pmatrix}; &
\makebox[0.07in]{}
{\bf c}_0=
\begin{pmatrix}0 \\ 0\end{pmatrix}\rightarrow
{\bf c}_1=
\begin{pmatrix}0 \\ 1\end{pmatrix}\rightarrow
{\bf c}_2=
\begin{pmatrix}0 \\ 0\end{pmatrix}; \\[0.15in]

{\bf c}_0=
\begin{pmatrix}0 \\ 3\end{pmatrix}\rightarrow
{\bf c}_1=
\begin{pmatrix}1 \\ 3\end{pmatrix}\rightarrow
{\bf c}_2=
\begin{pmatrix}0 \\ 3\end{pmatrix}; &
\makebox[0.07in]{}
{\bf c}_0=
\begin{pmatrix}3 \\ 0\end{pmatrix}\rightarrow
{\bf c}_1=
\begin{pmatrix}3 \\ 1\end{pmatrix}\rightarrow
{\bf c}_2=
\begin{pmatrix}3 \\ 0\end{pmatrix}; \\[0.15in]

{\bf c}_0=
\begin{pmatrix}0 \\ 3\end{pmatrix}\rightarrow
{\bf c}_1=
\begin{pmatrix}0 \\ 2\end{pmatrix}\rightarrow
{\bf c}_2=
\begin{pmatrix}0 \\ 3\end{pmatrix}; &
\makebox[0.07in]{}
{\bf c}_0=
\begin{pmatrix}3 \\ 0\end{pmatrix}\rightarrow
{\bf c}_1=
\begin{pmatrix}2 \\ 0\end{pmatrix}\rightarrow
{\bf c}_2=
\begin{pmatrix}3 \\ 0\end{pmatrix}; \\[0.15in]

{\bf c}_0=
\begin{pmatrix}3 \\ 3\end{pmatrix}\rightarrow
{\bf c}_1=
\begin{pmatrix}2 \\ 3\end{pmatrix}\rightarrow
{\bf c}_2=
\begin{pmatrix}3 \\ 3\end{pmatrix}; &
\makebox[0.07in]{}
{\bf c}_0=
\begin{pmatrix}3 \\ 3\end{pmatrix}\rightarrow
{\bf c}_1=
\begin{pmatrix}3 \\ 2\end{pmatrix}\rightarrow
{\bf c}_2=
\begin{pmatrix}3 \\ 3\end{pmatrix}.
\end{array}
\end{equation}
Note that the configuration sequences on each row of \eqref{eq:c_seq} is formed by permuting elements of the reduced configuration sequence on the corresponding row in \eqref{eq:red_c_seq}. Finally, each configuration sequence in \eqref{eq:c_seq} can be replaced with sequences of states, forming the set $\mathcal{W}_2$ which consists of the following sequences of states $\phi_0\rightarrow\phi_1\rightarrow\phi_2$ (Here $|$ separates the two ancilla registers and each block of 3 sequences corresponds to the configuration sequence in the associated row and position in \eqref{eq:c_seq}):
\begin{equation}\label{eq:seq}
\begin{array}{cc}
000|000 \rightarrow 100|000 \rightarrow 000|000 & \qquad 000|000 \rightarrow 000|100 \rightarrow 000|000 \\
000|000 \rightarrow 010|000 \rightarrow 000|000 & \qquad 000|000 \rightarrow 000|010 \rightarrow 000|000 \\
000|000 \rightarrow 001|000 \rightarrow 000|000 & \qquad 000|000 \rightarrow 000|001 \rightarrow 000|000 \\[0.1in]

000|111 \rightarrow 100|111 \rightarrow 000|111 & \qquad 111|000 \rightarrow 111|100 \rightarrow 111|000 \\
000|111 \rightarrow 010|111 \rightarrow 000|111 & \qquad 111|000 \rightarrow 111|010 \rightarrow 111|000 \\
000|111 \rightarrow 001|111 \rightarrow 000|111 & \qquad 111|000 \rightarrow 111|001 \rightarrow 111|000 \\[0.1in]

000|111 \rightarrow 000|011 \rightarrow 000|111 & \qquad 111|000 \rightarrow 011|000 \rightarrow 111|000 \\
000|111 \rightarrow 000|101 \rightarrow 000|111 & \qquad 111|000 \rightarrow 101|000 \rightarrow 111|000 \\
000|111 \rightarrow 000|110 \rightarrow 000|111 & \qquad 111|000 \rightarrow 110|000 \rightarrow 111|000 \\[0.1in]

111|111 \rightarrow 011|111 \rightarrow 111|111 & \qquad 111|111 \rightarrow 111|011 \rightarrow 111|111 \\
111|111 \rightarrow 101|111 \rightarrow 111|111 & \qquad 111|111 \rightarrow 111|101 \rightarrow 111|111 \\
111|111 \rightarrow 110|111 \rightarrow 111|111 & \qquad 111|111 \rightarrow 111|110 \rightarrow 111|111 \\
\end{array}
\end{equation}
We observe that the first (top left) block of three sequences in \eqref{eq:seq} sums up to a term $3\lambda_1^2\cdot\frac{1}{z-E_1}|000\rangle\langle 000|$, with each sequence contributing a term $\lambda_1\cdot\frac{1}{|z-E_1|}\cdot\lambda_1$ in the final upper bound (Equation \ref{eq:Tr_red_c0cr}). Similarly we see that the top right block of \eqref{eq:seq} sums up to $3\lambda_2^2\cdot\frac{1}{z-E_1}|000\rangle\langle 000|$. Recall that $\Pi_-$ is the projector onto the low energy subspace $\mathcal{L}_-=\mathcal{L}_-^{(1)}\otimes\mathcal{L}_-^{(2)}$ with each $\mathcal{L}_-^{(i)}=\text{span}\{|000\rangle,|111\rangle\}$. Adding up the terms in all the sequences in \eqref{eq:seq} gives a term $3(\lambda_1^2+\lambda_2^2)\cdot\frac{1}{z-E_1}\Pi_-$, which is symmetric with respect to the permutation of registers.

Also observe that we are able to calculate the coefficient $3(\lambda_1^2+\lambda_2^2)\cdot\frac{1}{z-E_1}$ in the sum over \emph{all} blocks of sequences in \eqref{eq:seq} by {only inspecting the first row}, gleaning two terms with coefficients $3\lambda_1^2\cdot\frac{1}{z-E_1}$ and $3\lambda_2^2\cdot\frac{1}{z-E_1}$. This is because the set $\mathcal{W}_r$ is invariant with respect to the operation of flipping all the bits of any set of registers (recall discussion prior to Equation \ref{eq:Tr_zero_phir}). Examining \ref{eq:seq} one could find that for instance flipping all the bits in the first register of the top left block yields the block on the right of third row. Flipping all the bits in the second register of the top left block yields the block on the left of the second row. Flipping all the bits in both registers yields the bottom right block, etc. Therefore in order to find the coefficients to $T_2$ it suffices to focus on only the sequences $\phi_0\rightarrow\phi_1\rightarrow\phi_2$ where $\phi_0=000|000$ (Equation \ref{eq:Tr_zero_phir}). 

\end{document}